\documentstyle[aps,multicol,epsf,rotate]{revtex}

\begin{document}

\draft

\title{Scaling of the distribution of price fluctuations of \\ 
individual companies}

\author{Vasiliki Plerou$^{1,2}$, Parameswaran Gopikrishnan$^{1}$, 
        Lu\'{\i}s A.\ Nunes Amaral$^{1}$, \\ Martin Meyer$^{1}$, and
        H. Eugene Stanley$^{1}$}

\address{ $^{1}$ Center for Polymer Studies and Dept. of Physics, Boston
        University, Boston, MA 02215, USA \\ $^{2}$Department of
        Physics, Boston College, Chestnut Hill, MA
        02167, USA }

\date{Last modified: \today.  Printed: \today}

\maketitle

\begin{abstract}

        We present a phenomenological study of stock price
        fluctuations of individual companies. We systematically
        analyze two different databases covering securities from the
        three major US stock markets: (a) the New York Stock Exchange,
        (b) the American Stock Exchange, and (c) the National
        Association of Securities Dealers Automated Quotation stock
        market. Specifically, we consider (i) the trades and quotes
        database, for which we analyze 40 million records for 1000 US
        companies for the 2-year period 1994--95, and (ii) the Center
        for Research and Security Prices database, for which we
        analyze 35 million daily records for approximately 16,000
        companies in the 35-year period 1962--96. We study the
        probability distribution of returns over varying time scales
        $\Delta t$, where $\Delta t$ varies by a factor of $\approx
        10^5$---from 5$\,$min up to $\approx$ 4 $\,$years. For time
        scales from 5~min up to approximately 16~days, we find that
        the tails of the distributions can be well described by a
        power-law decay, characterized by an exponent $\alpha \approx
        3$ ---well outside the stable L\'evy regime $0 < \alpha <
        2$. For time scales $\Delta t \gg (\Delta t)_{\times} \approx
        16\,$days, we observe results consistent with a slow
        convergence to Gaussian behavior. We also analyze the role of
        cross correlations between the returns of different companies
        and relate these correlations to the distribution of returns
        for market indices.

\end{abstract}

\begin{multicols}{2}

%%%%%%%%%%%%%%%%%%%%%%%%%%%%% SECTION
\section{Introduction}
%%%%%%%%%%%%%%%%%%%%%%%%%%%%%%%%%%%%%%%%%%%%%%%%%%%%%%%%%

The study of financial markets poses many challenging questions.  For
example, how can one understand a strongly fluctuating system that is
constantly driven by external information?  And, how can one account
for the role of the feedback between the markets and the outside
world, or of the complex interactions between traders and assets?  An
advantage for the researcher trying to answer these questions is the
availability of huge amounts of data for analysis.  Indeed, the
activities at financial markets result in several observables, such as
the values of different market indices, the prices of the different
stocks, trading volumes, etc.

Some of the most widely studied market observables are the values of
market indices. Previous empirical
studies~\cite{Bouchaud98,ms,Kondor98,palermo,kent,Mandelbrot63,%%%
Mantegna95,Ghasghaie96,Arneodo98,Ausloos,Dietrich,Pagan96} show that
the distribution of fluctuations ---measured by the returns--- of
market indices has slow decaying tails and that the distributions
apparently retain the same functional form for a range of time
scales~\cite{Bouchaud98,ms,Mandelbrot63,Mantegna95}.  Fluctuations in
market indices reflect average behavior of the price fluctuations of
the companies comprising them.  For example, the S\&P 500 is defined
as the sum of the market capitalizations (stock price multiplied by
the number of outstanding shares) of 500 companies representative of
the US economy.

Here, we focus on a more ``microscopic'' quantity: individual
companies. We analyze the tic-by-tic data~\cite{tbt} for the 1000
publicly-traded US companies with the largest market capitalizations
and systematically study the statistical properties of their stock
price fluctuations.  A preliminary study~\cite{Gopi98} reported that
the distribution of the 5$\,$min returns for 1000 individual companies
and the S\&P 500 index decays as a power-law with an exponent $\alpha
\approx 3$ ---well outside the stable L\'evy regime ($\alpha <
2$). Earlier independent studies on individual stock returns on longer
time scales yield similar results~\cite{Lux96}. These findings raise
the following questions:

First, how does the nature of the distribution of individual stock
returns change with increasing time scale $\Delta t$? In other words,
does the distribution retain its power-law functional form for longer
time scales, or does it converge to a Gaussian, as found for market
indices~\cite{Mantegna95,Gopi99}? If the distribution indeed converges
to Gaussian behavior, how fast does this convergence occur? For the
S\&P 500 index, for example, one finds the distribution of returns to
be consistent with a {\it non-stable} power-law functional form
($\alpha \approx 3$) until approximately 4~days, after which an onset
of convergence to Gaussian behavior is found~\cite{Gopi99}.

Second, why is it that the distribution of returns for individual
companies and for the S\&P 500 index have the same asymptotic form?  This
finding is unexpected, since the returns of the S\&P 500 are the
weighted sums of the returns of 500 companies.  Hence, we would
expect the S\&P 500 returns to be distributed approximately as a
Gaussian, unless there were significant dependencies between the
returns of different companies which prevent the central limit theorem
from applying.

To answer the first question, we extend previous work\cite{Gopi98} on
the distribution of returns for 5 min returns by performing empirical
analysis of individual company returns for time scales up to
46~months. Our analysis uses two distinct data-bases detailed
below. We find that the cumulative distribution of individual-company
returns is consistent with a power-law asymptotic behavior with
exponent $\alpha \approx 3$, which is outside the stable L\'evy
regime. We also find that these distributions appear to retain the
same functional form for time scales up to approximately 16~days. For
longer time scales, we observe results consistent with a slow
convergence to Gaussian behavior.

To answer the second question, we randomize each of the 500 time
series of returns for the constituent 500 stocks of the S\&P 500
index. A surrogate ``index return'' thus constructed from the
randomized time series, shows fast convergence to Gaussian. Further,
we find that the functional form of the distribution of returns
remains unchanged for different system sizes (measured by the market
capitalization) while the standard deviation decays as a power-law of
market capitalization. 

The organization of this paper is as follows. Section II describes the
databases studied and the data analyzed. Sections III, IV, and V
present results for the distribution of returns for individual
companies for a wide range of time scales. Section VI discusses the
role of cross-correlations between companies and possible reasons why
market indices have statistical properties very similar to those of
individual companies. Section VII contains some concluding remarks.

%%%%%%%%%%%%%%%%%%%%%%%%%%%%% SECTION
\section{The Data analyzed}
%%%%%%%%%%%%%%%%%%%%%%%%%%%%%%%%%%%%%%%%%%%%%%%%%%%%%%%%%%%%%%

We analyze two different databases covering securities from the
three major US stock markets, namely (i) the New York Stock Exchange
(NYSE), (ii) the American Stock Exchange (AMEX), and (iii) the
National Association of Securities Dealers Automated Quotation
(Nasdaq) stock market.  NYSE is the oldest stock exchange, tracing its
origin to the Buttonwood Agreement of 1792\cite{webNYSE}. The NYSE is
an agency auction market, that is, trading at the NYSE takes place by
open bids and offers by Exchange members, acting as agents for
institutions or individual investors. Buy and sell orders are brought
to the trading floor, and prices are determined by the interplay of
supply and demand.  As of the end of November 1998, the NYSE lists
over 3,100 companies. These companies have over $2 \times 10^{11}$
shares, worth approximately USD~$ 10^{13} $, available for trading on
the Exchange.

In contrast to NYSE, Nasdaq uses computers and telecommunications
networks which create an electronic trading system wherein the market
participants meet over the computer rather than face to face. Nasdaq's
share volume reached $1.6\times 10^{11}$ shares in 1997 and dollar
volume reached USD~$4.4\times 10^{12}$.  As of December 1998, the
Nasdaq Stock Market listed over 5,400 US and non-US
companies\cite{webNasdaq}. Nasdaq and AMEX, have merged on October
1998, after the end of the period studied in this work.

The first database we consider is the trades and quotes (TAQ)
database\cite{webTAQ}, for which we analyze the 2-year period 
January 1994 to  December 1995.  The TAQ database, which is
published by NYSE since 1993, covers {\it all\/} trades at the three
major US stock markets.  This huge database is available in the form
of CD-ROMs.  The rate of publication was 1 CD-ROM per month for the
period studied, but recently has increased to 2--3 CD-ROMs per
month. The total number of transactions for the largest 1000 stocks is
of the order of $10^9$ in the 2-year period studied.

The second database we analyze is the Center for Research and Security
Prices (CRSP) database\cite{webCRSP}.  The CRSP Stock Files cover
common stocks listed on NYSE beginning in 1925, the AMEX beginning in
1962, and the Nasdaq Stock Market beginning in 1972. The files provide
complete historical descriptive information and market data including
comprehensive distribution information, high, low and closing prices,
trading volumes, shares outstanding, and total returns\cite{crspts}.

The CRSP Stock Files provide monthly data for NYSE beginning December
1925 and daily data beginning July 1962.  For the AMEX, both monthly
and daily data begin in July 1962.  For the Nasdaq Stock Market, both
monthly and daily data begin in July 1972.

We also analyze the S\&P 500 index, which comprises 500 companies
chosen for market size, liquidity, and industry group representation
in the US.  In our study, we analyze data with a recording frequency
of less than 1~min that cover the 13 years from January 1984 to
December 1996. The total number of data points in this 13-year period
exceeds $4.5 \times 10^6$.

%%%%%%%%%%%%%%%%%%%%%%%%%%%%% SECTION
\section{The distribution of returns for $\Delta t < 1$ day }
%%%%%%%%%%%%%%%%%%%%%%%%%%%%%%%%%%%%%%%%%%%%%%%%%%%%%%%

The basic quantity studied for individual companies --- $i=1,2,\dots
,1000$ --- is the market capitalization $S_i(t)$, defined as the share
price multiplied by the number of outstanding shares.  The time $t$
runs over the working hours of the stock exchange---removing nights,
weekends and holidays\cite{work_hrs}. For each company, we analyze the
return

\begin{equation}
G_i\equiv G_i(t,\Delta t)\equiv \ln S_i(t+\Delta t) - \ln S_i(t)\,. 
\label{return}
\end{equation}
For small changes in $S_i(t)$, the return $ G_i(t,\Delta t)$ is
approximately the forward relative change,
\begin{equation}
G_i(t, \Delta t) \approx {S_i(t+\Delta t)-S_i(t)\over S_i(t)}.
\label{relChange}
\end{equation}
For time scales shorter than 1~day, we analyze the data from the TAQ
database.  We consider the largest 1000 companies\cite{lar1000}, in
decreasing order of values of their market capitalization on the first
trading day, 3 January 1994. We sample the price of these 1000
companies at 5$\,$min intervals\cite{stime}. In order to obtain time
series for market capitalization, we multiply the stock price of each
company by the number of outstanding shares for that company at each
sampling time.  We thereby generate a time series, sampled at 5~min
intervals, for the market capitalizations of each of the largest 1000
companies. Each of the 1000 time series has approximately 40,000 data
points---corresponding to the number of 5~min intervals in the 2-year
period---or about 40 million data points in total. For each time series
of market capitalizations, we compute the 5~min returns using
Eq.~(\ref{return}). We filter the data to remove spurious events, such
as occur due to the inevitable recording errors\cite{filter}.

\subsection{The distribution of returns for $\Delta t = 5$~min}
%%%%%%%%%%%%%%%%%%%%%%%%%%%%%%%%%

\begin{figure}
\narrowtext
\centerline{
\epsfysize=0.5\columnwidth{\rotate[r]{\epsfbox{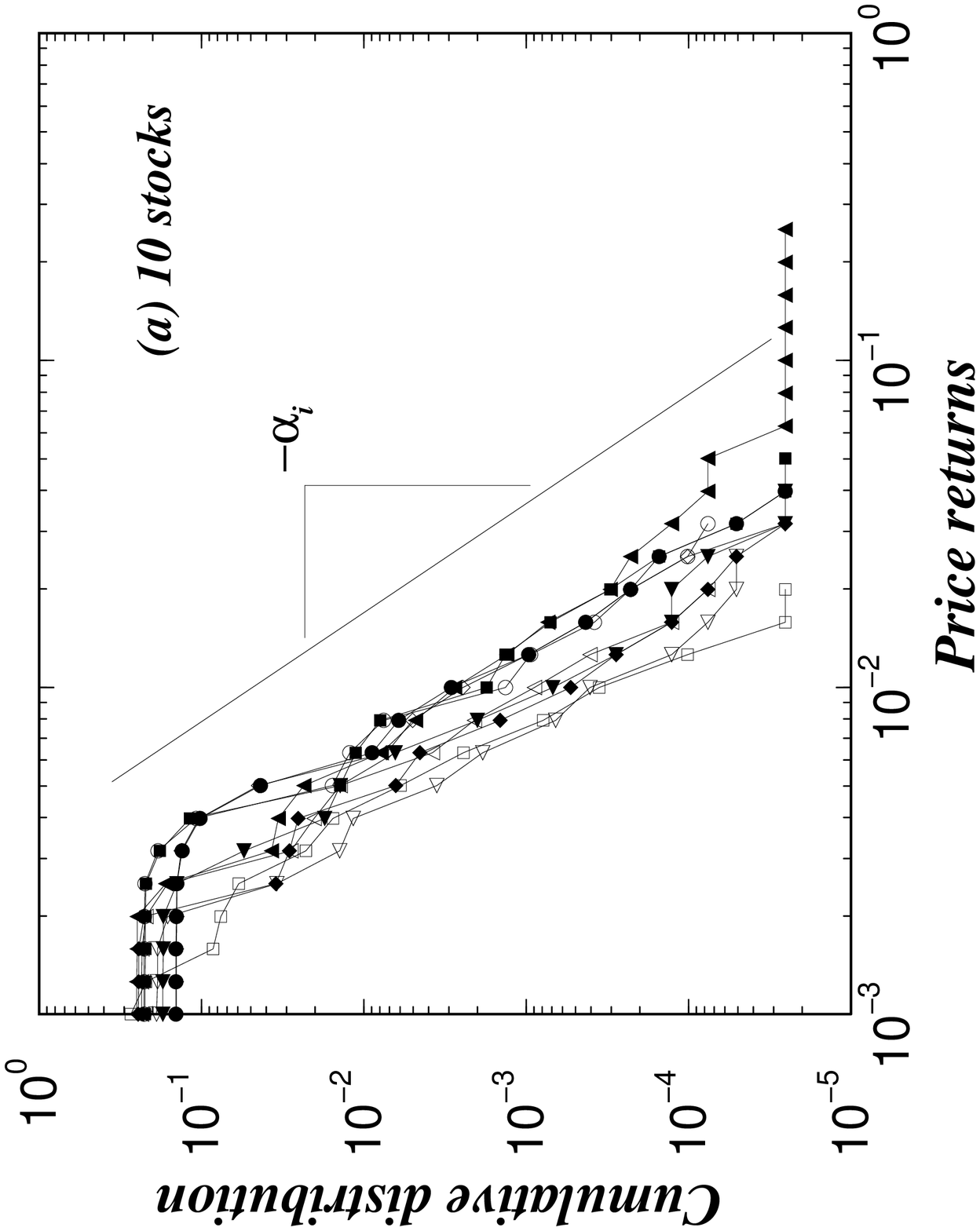}}}
\hspace*{0.5cm}
\epsfysize=0.5\columnwidth{\rotate[r]{\epsfbox{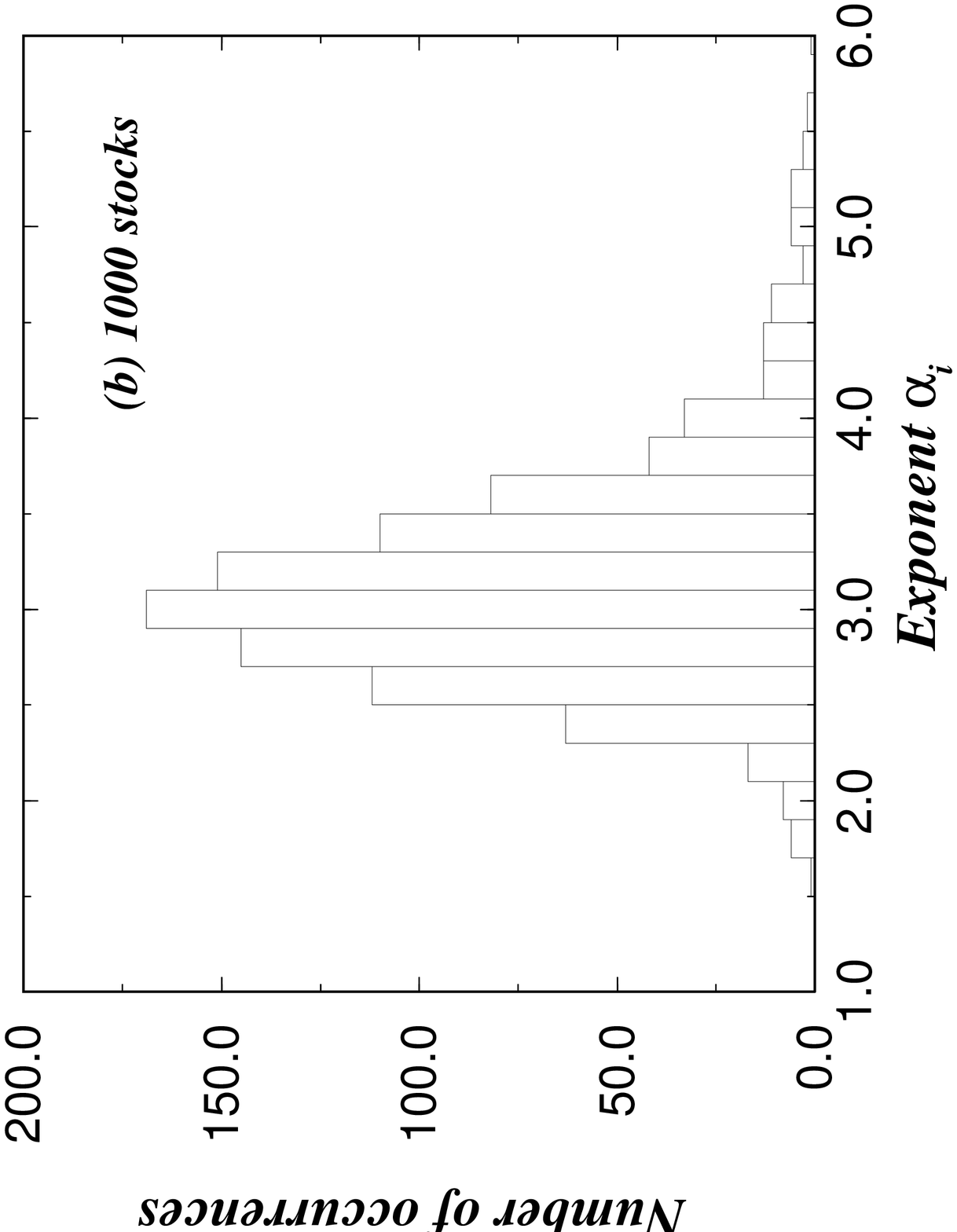}}}
}
%\vspace*{0.5cm}
\centerline{
\epsfysize=0.5\columnwidth{\rotate[r]{\epsfbox{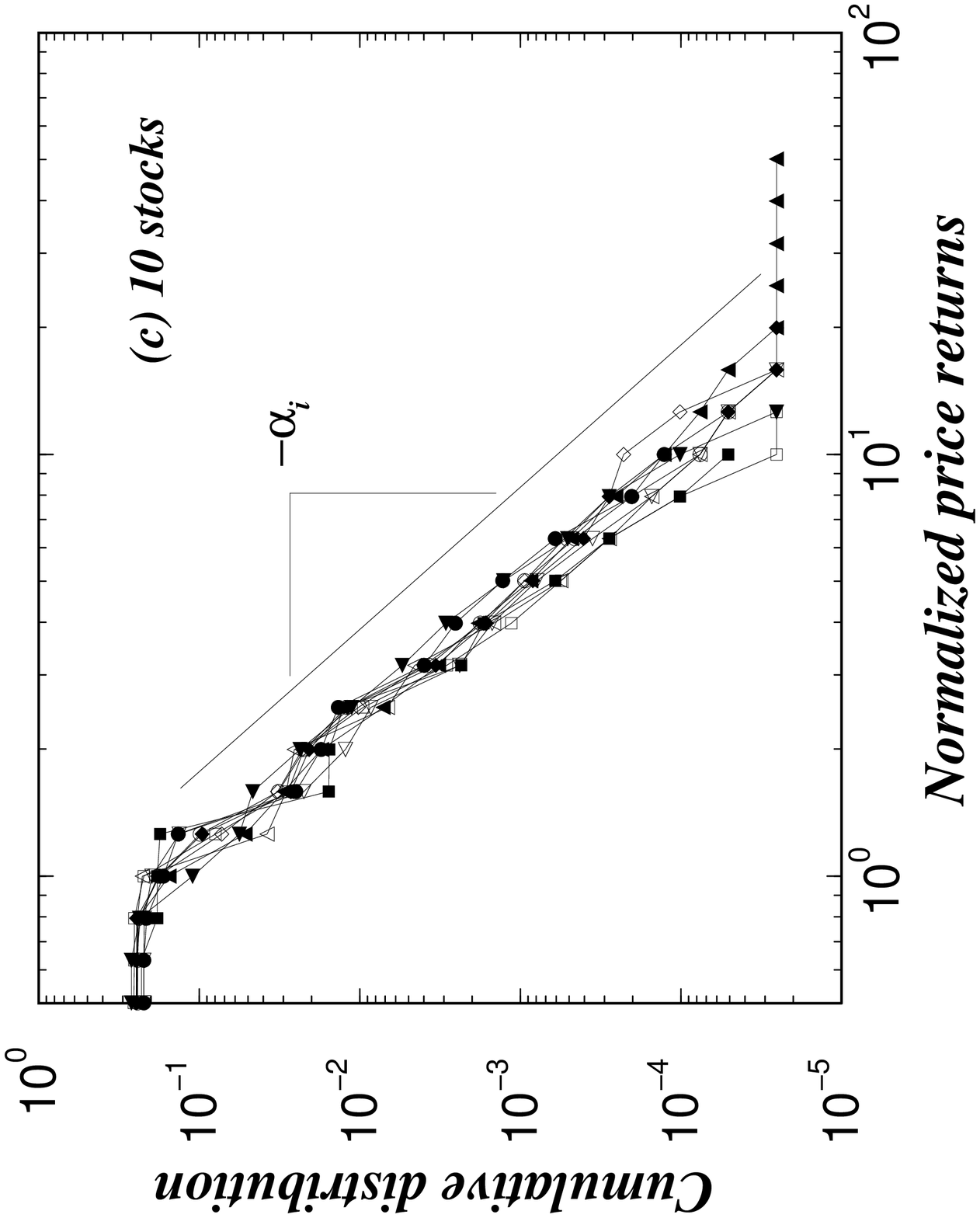}}}
}
\caption{ {\it (a)} Cumulative distributions $P(g>x)$ for the positive
tails of 10 randomly-selected companies. Note that they are all
consistent with a power-law asymptotic behavior. {\it (b)} The
histogram of the power-law exponents obtained by power-law regression
fits to the individual cumulative distribution functions, where the
fit is for all $x$ larger than 2 standard deviations. Note that this
histogram is not normalized---the y-axis indicates the number of
occurrences of the exponent. {\it (c)} Cumulative distributions of
the 10 randomly chosen companies in (a) scaled by the standard
deviation calculated from the entire 2-year period.}
\label{taq}
\end{figure}

Figure~\ref{taq}(a) shows the cumulative distributions of returns
$G_i$ for $\Delta t = 5$~min --- the probability of a return larger
than or equal to a threshold --- for 10 individual companies randomly
selected from the 1000 companies that we analyze. For each company
$i$, the asymptotic behavior of the functional form of the cumulative
distribution is ``visually'' consistent with a power-law,
\begin{equation}
P(G_i>x)\sim{1\over x^{\alpha_i}} \,,
\label{def_alphai}
\end{equation}
where $\alpha_i$ is the exponent characterizing the power-law decay.  In
Fig.~\ref{taq}(b) we show the histogram for $\alpha_i$, obtained from
power-law regression-fits to the positive tails of the individual cumulative
distributions of all 1000 companies studied. The histogram has most probable
value $\alpha_{MP} = 3 $.

\begin{figure}
\narrowtext
%\vspace*{0.5cm}
\centerline{
\epsfysize=0.5\columnwidth{\rotate[r]{\epsfbox{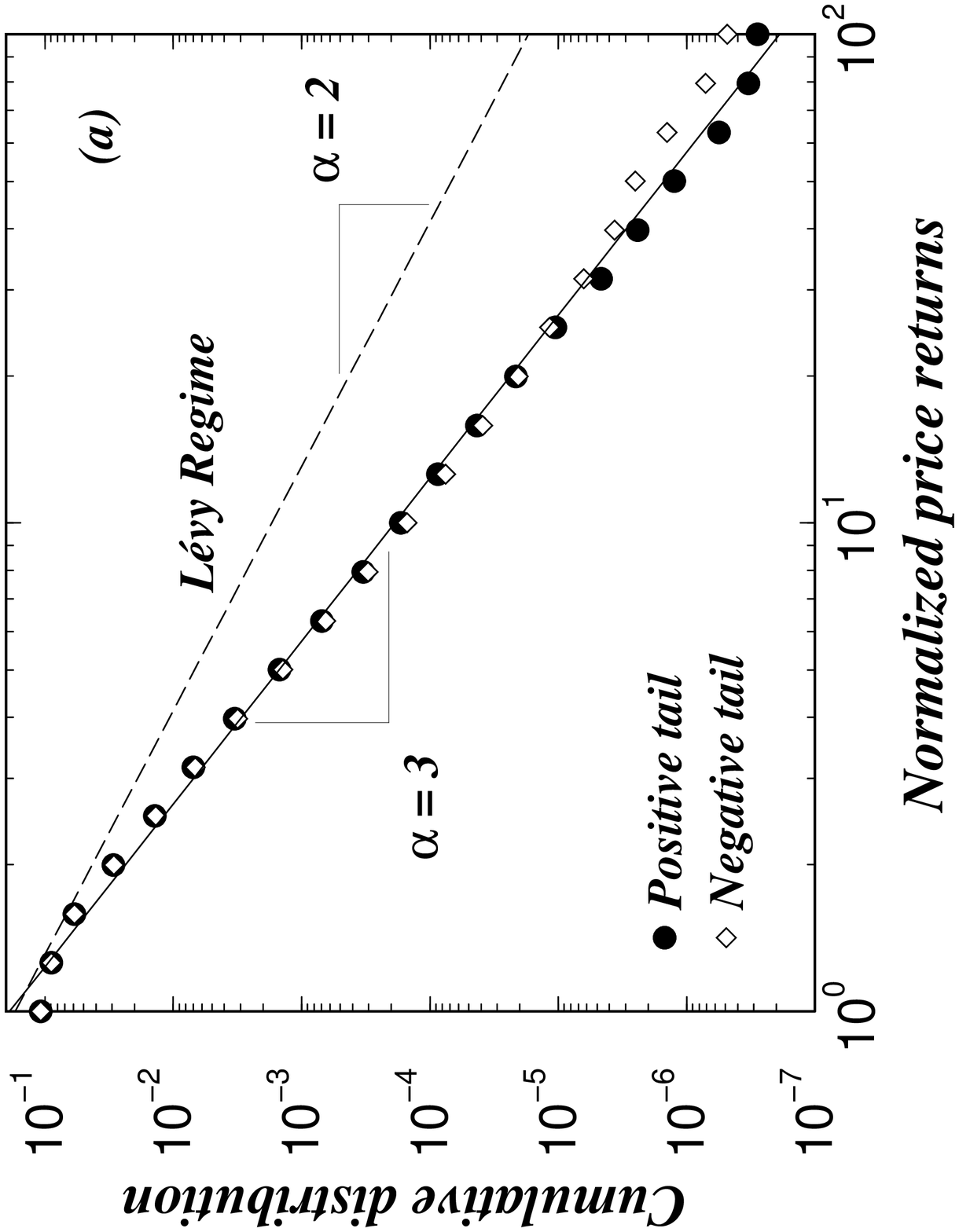}}}
\hspace*{0.5cm}
\epsfysize=0.5\columnwidth{\rotate[r]{\epsfbox{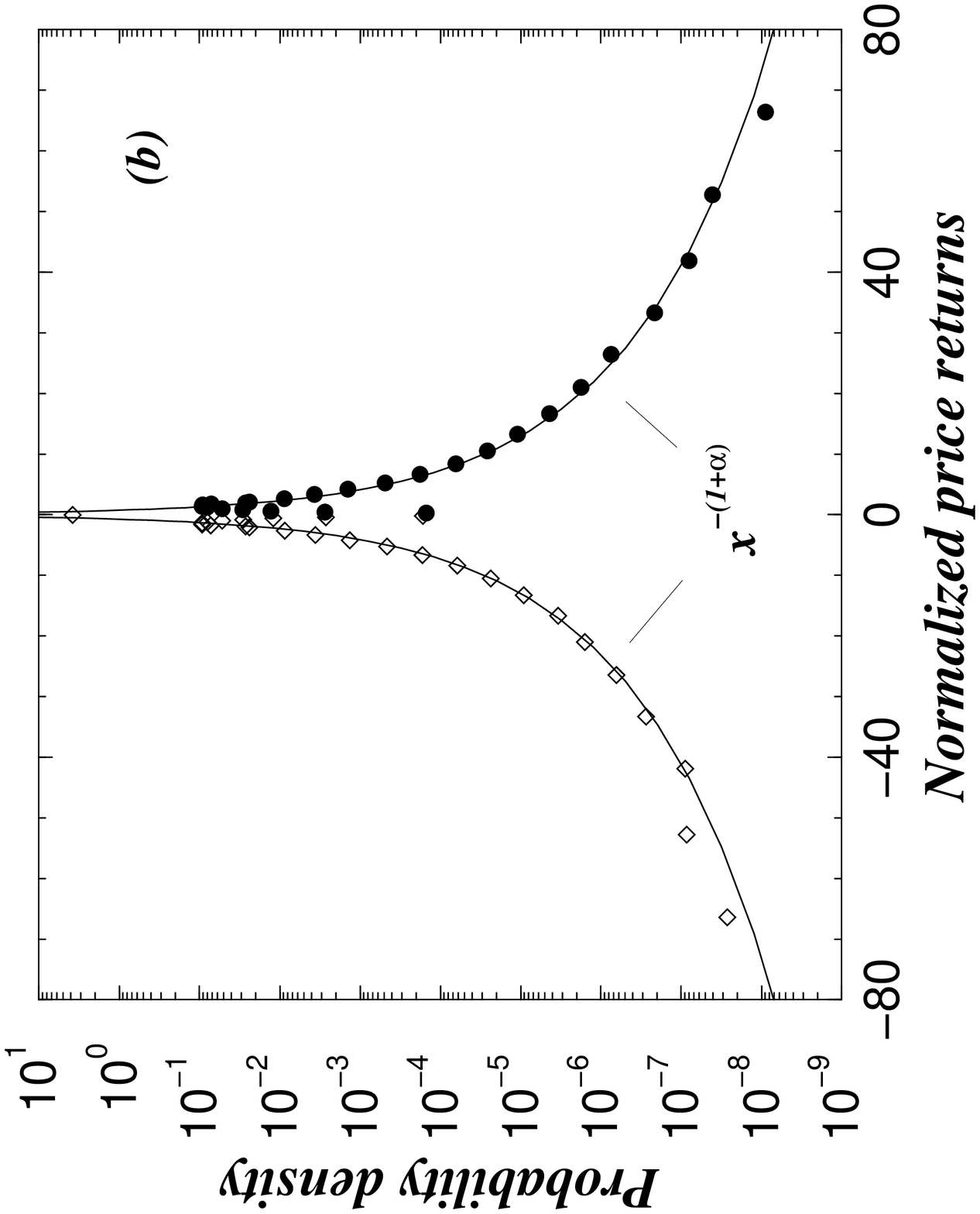}}} 
}
\vspace*{0.5cm}
\caption{ {\it (a)} Cumulative distributions of the positive and
negative tails of the normalized returns of the 1000 largest companies
in the TAQ database for the 2-year period 1994--1995. The solid line
is a power-law regression fit in the region $2\leq x \leq 80$. {\it
(b)} Probability density function of the normalized returns. The
values in the center of the distribution arise from the discreteness
in stock prices, which are set in units of fractions of USD, usually
1/8, 1/16, or 1/32. The solid curve is a power-law fit in the region
$2\leq x \leq 80$.  We find $\alpha = 3.10\pm 0.03$ for the positive
tail and $\alpha = 2.84 \pm 0.12$ for the negative tail. }
\label{sctaq}
\end{figure}

Next, we compute the time-averaged volatility $v_i \equiv v_i(\Delta
t)$ of company $i$ as the standard deviation of the returns over the
2-year period
\begin{equation}
{v_i}^2\equiv \langle {G_i}^2 \rangle_T -{\langle G_i \rangle_T}^2\,,
\label{eq_defv}
\end{equation}
where $\langle\dots\rangle_T$ denotes a time average over the 40,000
data points of each time series, for the 2-year period studied.
Figure~\ref{taq}(a) suggests that the widths of the individual
distributions differ for different companies; indeed, companies with small
values of market capitalization are likely to fluctuate more. In order
to compare the returns of different companies with different
volatilities, we define the normalized return $g_i \equiv g_i(t,\Delta
t)$ as
\begin{equation}
g_i \equiv {G_i -\langle G_i \rangle_T \over v_i}\,.
\label{eq_defg}
\end{equation}
Figure~\ref{taq}(c) shows the ten cumulative distributions of the
normalized returns $ g_i$ for the same ten companies as in
Fig~\ref{taq}(a). The distributions for all 1000 normalized returns
$g_i$ have similar functional forms to these ten. Hence, to obtain
better statistics, we compute a {\it single} distribution of all the
normalized returns. The cumulative distribution $P(g>x)$ shows a
power-law decay [Fig~\ref{sctaq}(a)],
\begin{equation}
P(g>x) \sim {1\over x^{\alpha}}\,.
\label{def_alpha}
\end{equation} 
Regression fits in the region $2 \leq g\leq 80$ yield 
\begin{equation}
\alpha = \cases{ 3.10 \pm 0.03 & (positive tail) \cr
2.84 \pm 0.12 & (negative tail)}.
\label{alpha_val}
\end{equation}
These estimates~\cite{errorbars} of the exponent $\alpha$ are well outside
the stable L\'evy range, which requires $0< \alpha < 2$.

\begin{figure}
\narrowtext
\centerline{
\epsfysize=0.6\columnwidth{\rotate[r]{\epsfbox{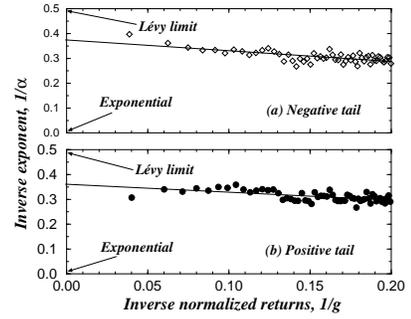}}} 
}
\vspace{1cm}
\caption{ The inverse local slope of $P(g)$, $\zeta^{-1}(g)\equiv
-\left(d \log P(g) / d \log g\right)$ as a function of the inverse
normalized returns $1/g$ for {\it (a)} the negative tail and {\it (b)}
the positive tail\protect\cite{Gopi99,Hill75}.  Each data point shown
is an average over 1000 events and the lines are linear regression
fits to the data. The linear regression fit over the range $0 \leq g
\leq 0.2 $ yields the values of the inverse asymptotic slopes,
$1/\alpha$; we find, $\alpha=2.84\pm0.12$ for the positive and
$\alpha=2.73\pm0.13$ for the negative tail. Note that the average over
all events used would be identical to the estimator for the asymptotic
slope proposed by Hill \protect\cite{Hill75}. }
\label{hilltaq}
\end{figure}

In order to obtain an alternative estimate for $\alpha$, we use the
methods of Hill~\cite{Pagan96,Gopi98,Lux96,Gopi99,Hill75}. We first
calculate the inverse of the local logarithmic slope of $P(g)$,
$\zeta^{-1}(g)\equiv\,d\log P(g)/d\log g$, where $g$ is
rank-ordered. We then estimate the asymptotic slope $\alpha$ by
extrapolating $\zeta$ as a function of $1/g \to
0$. Figure~\ref{hilltaq} shows the results for the negative and
positive tails, for the 5$\,$min returns for individual companies,
each using all returns larger than 5 standard deviations.
Extrapolation of the linear regression lines yield:
\begin{equation}
\alpha = \cases{ 2.84 \pm 0.12 & (positive tail) \cr
2.73 \pm 0.13 & (negative tail)}.
\end{equation} 

\subsection{Scaling of the distribution of returns for $\Delta t \le 
1\,$day}
%%%%%%%%%%%%%%%%%%%%%%%%%%%

The next logical step would be to extend the previous procedure to
time scales longer than 5~min.  However, this approach leads to
unreliable results, the reason being that the estimate of the time
averaged volatility---used to define the normalized returns of
Eq.~(\ref{eq_defg})---has estimation errors that increase with $\Delta
t$.  For the distribution of 5~min returns, the previous procedure
relies on 40,000 data points per company for the estimation of the
time averaged volatility.  For 500~min returns the number of data
points available is reduced to 400 per company which leads to a much
larger error in the estimate of $v_i(\Delta t)$.

\begin{figure}
\narrowtext
\centerline{
\epsfysize=0.5\columnwidth{\rotate[r]{\epsfbox{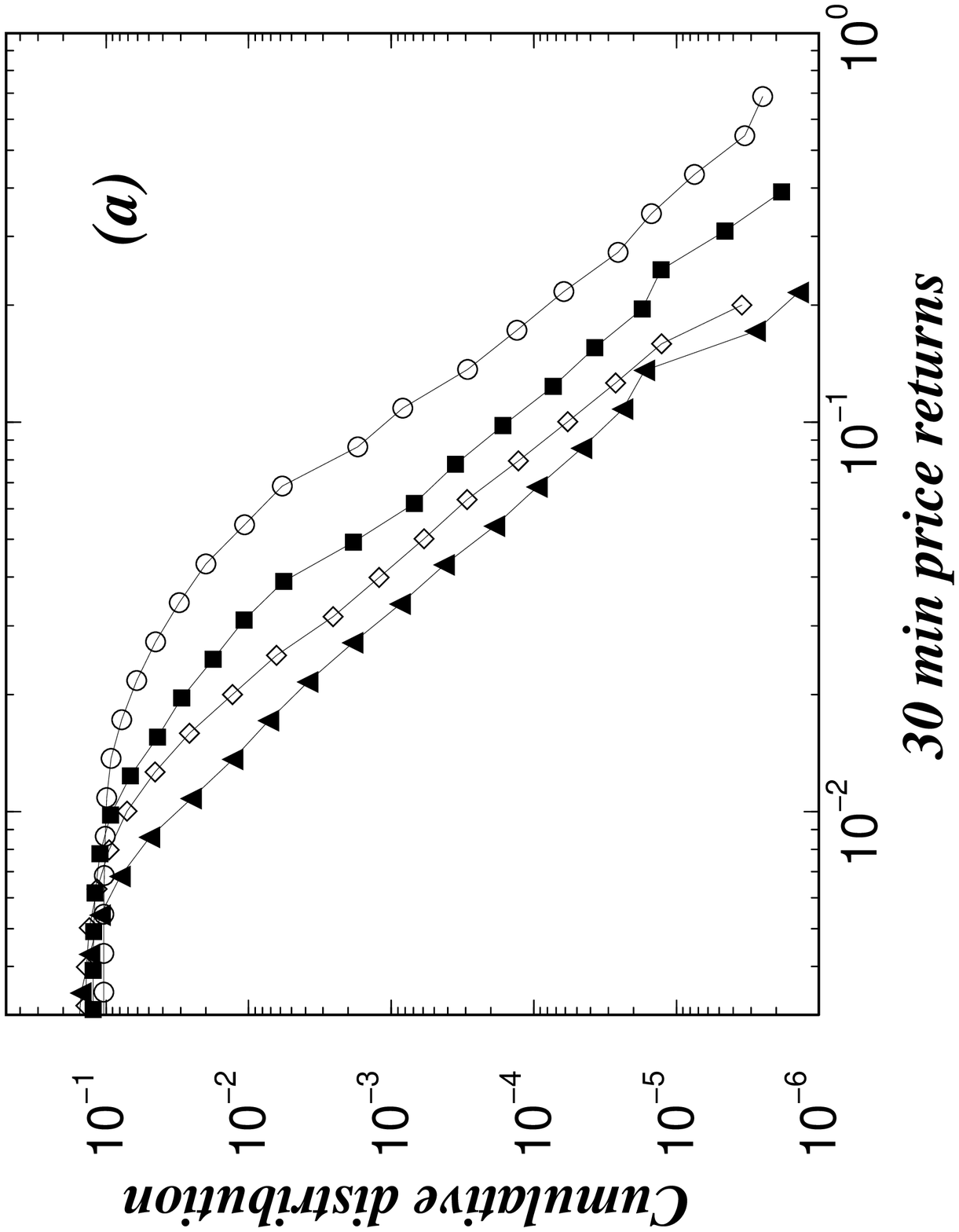}}}
\hspace*{0.5cm}
\epsfysize=0.5\columnwidth{\rotate[r]{\epsfbox{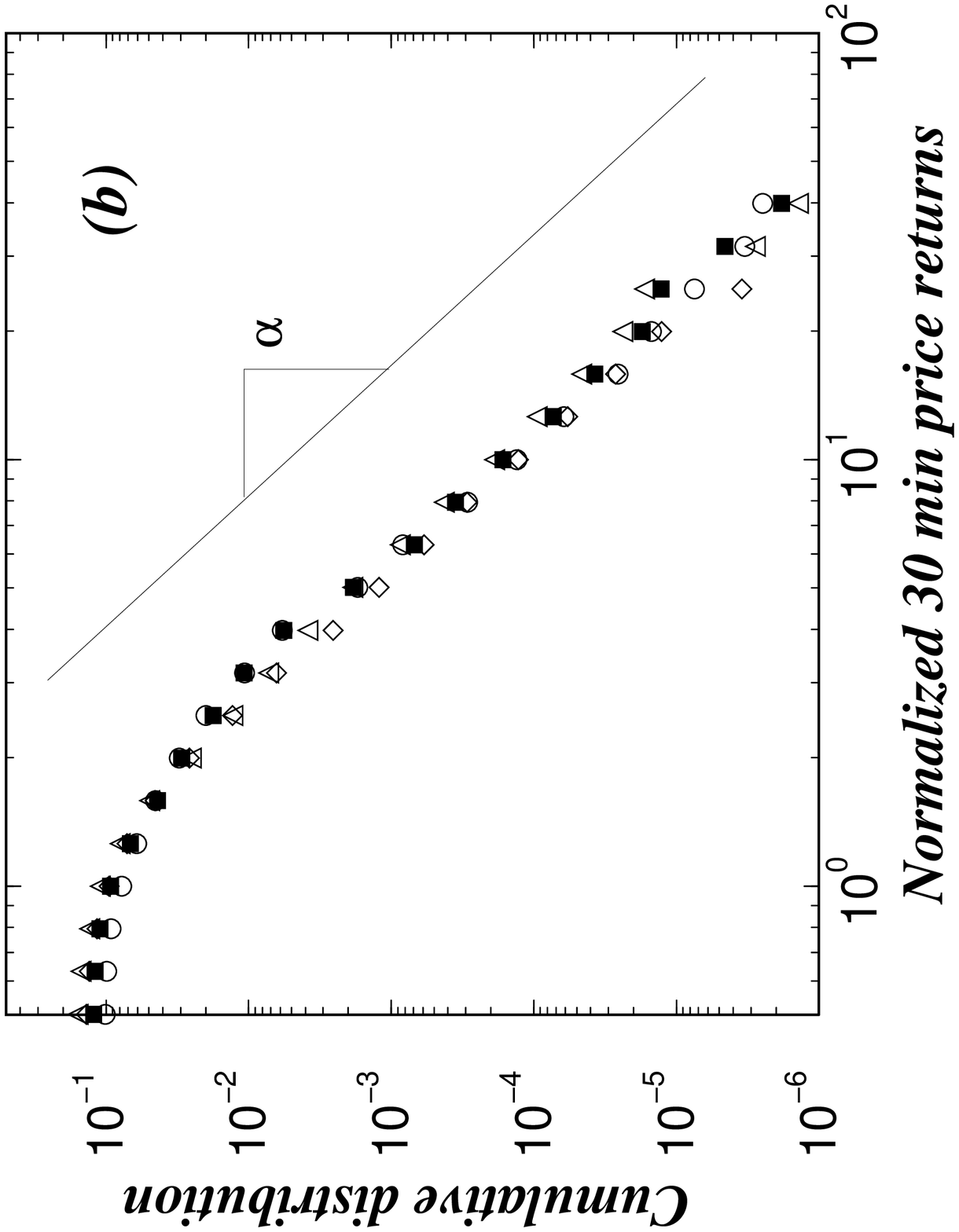}}}
}
\vspace{0.5cm}
\caption{ {\it (a)} Cumulative distribution of the conditional
probability $P(g > x |S)$ of the 30~min returns, for companies with
market capitalization $S$, from the TAQ database.  We define uniformly
spaced bins on a logarithmic scale.  We show the distribution of
returns for the 4 bins, $10^{9.8} < S \leq 10^{10.2}$, $10^{10.2} < S
\leq 10^{10.4}$, $10^{10.4} < S \leq 10^{10.6}$, and $10^{10.6} < S
\leq 10^{10.8}$. {\it (b)} Cumulative conditional distributions of
returns normalized by the average volatility $v_S(\Delta t)$ of each
bin.  Note that we find the same functional form for the different
values of $S$. }
\label{cond_taq}
\end{figure}

To circumvent the difficulty arising from the large uncertainty in
$v_i(\Delta t)$, we use an alternative procedure for estimating the
volatility\cite{Stanley96,Lee98,Amaral98} which relies on two
observations.  The first is that volatility decreases with market
capitalization [Fig.~\ref{varsize}].  The second is that companies with
similar market capitalization typically have similar volatilities.
Based on these observations, we make the hypothesis that the market
capitalization is the most influential factor in determining the
volatility,
\begin{equation}
v_i = v_i(S, \Delta t)\,.
\end{equation}
Hence, we group the returns of all the companies into ``bins''
according to the market capitalization of each company at the
beginning of the interval for which the return is computed.  We then
compute the conditional probability of the $\Delta t$ returns for each
of the bins of market capitalization. We define $G_S \equiv G_S(t,
\Delta t)$ as the $\Delta t$ returns of the subset of all companies
with market capitalization $S$, and we then calculate the cumulative
conditional probability $P(G_S \geq x | S)$.  Figure~\ref{cond_taq}(a)
shows $P(G_S \geq x | S)$ for 30~min returns for four different bins
of $S$. The functional form for each of each of the four distributions
is consistent with a power-law.

We define a normalized return
\begin{equation}
g_S \equiv g_S(t,\Delta t)\equiv{ G_S(\Delta t) -\langle G_S(\Delta
t)\rangle_S\over v_S(\Delta t)}\,,
\label{eq_g_tcond}
\end{equation}
where $\langle \cdots \rangle_S$ denotes an average over all returns of
all companies with market capitalization $S$. The average volatility $
v_S \equiv v_S (\Delta t)$ is defined through the relation,
\begin{equation}
{v_S}^2\equiv\langle {G_S}^2 \rangle_S -{\langle G_S \rangle_S}^2\,.
\label{eq_v_tcond}
\end{equation}

We show in Fig.~\ref{cond_taq}(b) the cumulative conditional
probability of the normalized 30~min returns $P(g_S \geq x | S)$ for
the same four bins shown in Fig.~\ref{cond_taq}(a). Visually, it seems
clear that these distributions have power-law functional forms with
similar values of $\alpha$. Hence, to obtain better statistics, we
consider the normalized returns for all values of $S$ and compute a
{\it single} cumulative distribution.

\begin{figure}
\narrowtext
\centerline{
\epsfysize=0.5\columnwidth{\rotate[r]{\epsfbox{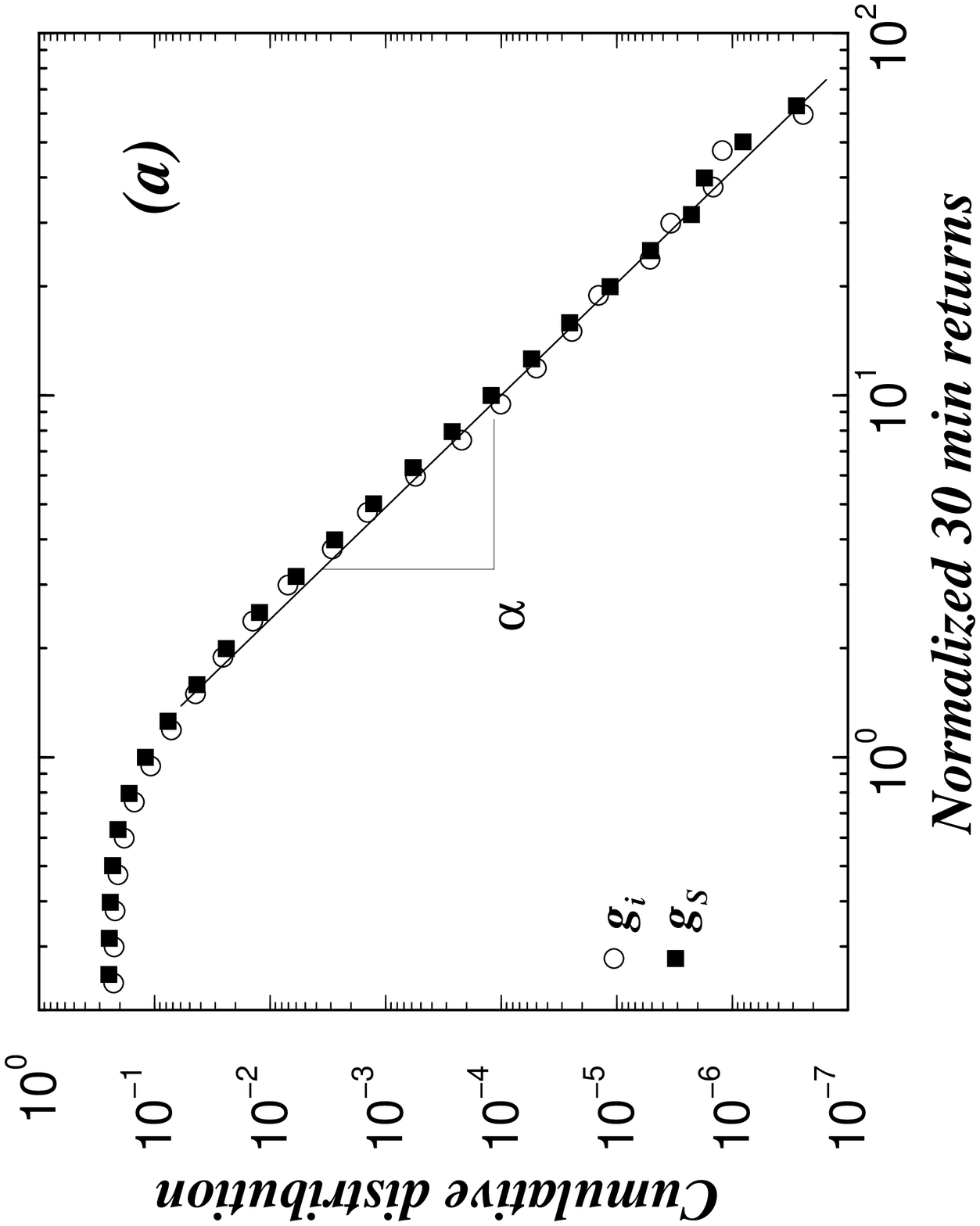}}}
\hspace{0.5cm}
\epsfysize=0.5\columnwidth{\rotate[r]{\epsfbox{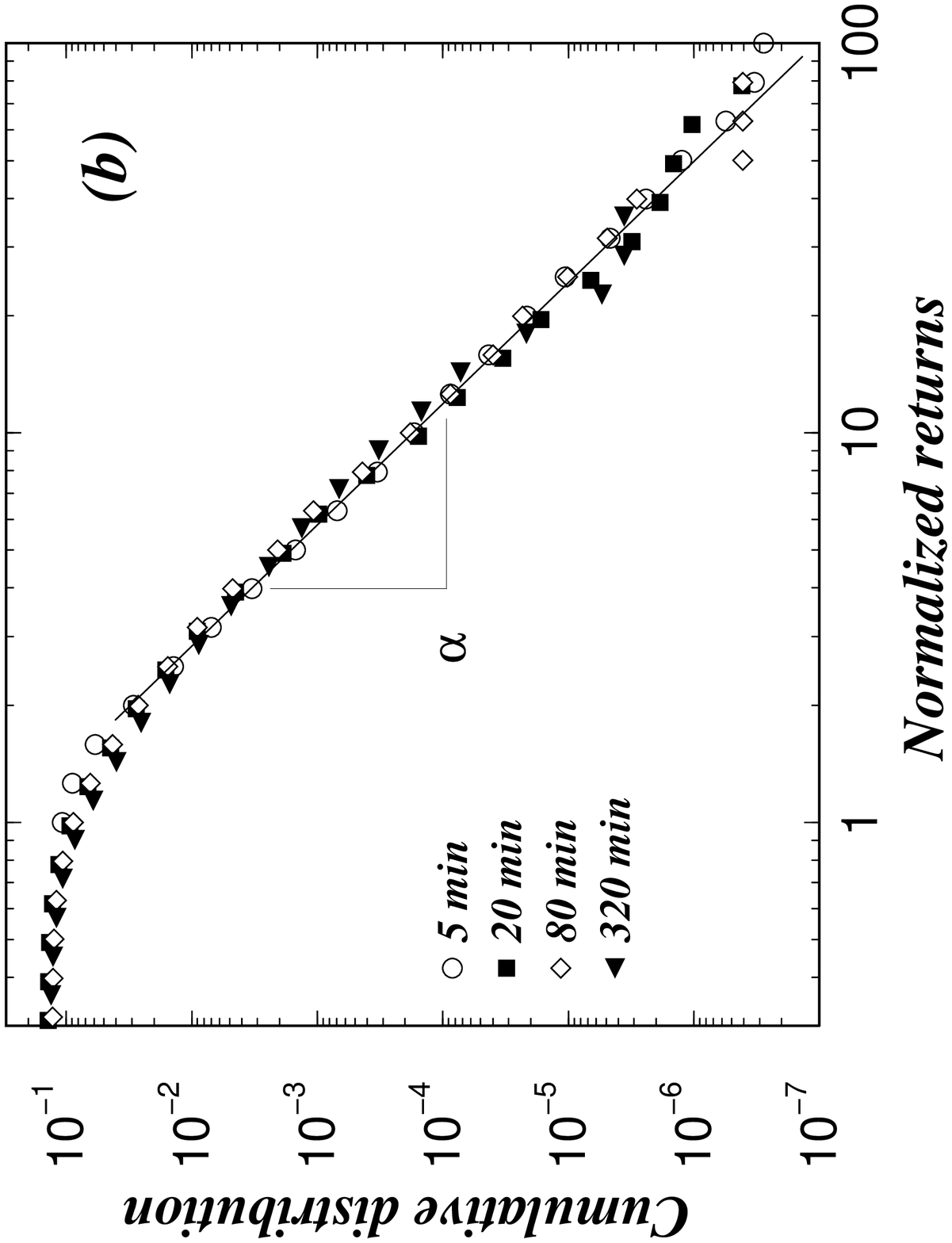}}}
}
\centerline{
\epsfysize=0.5\columnwidth{\rotate[r]{\epsfbox{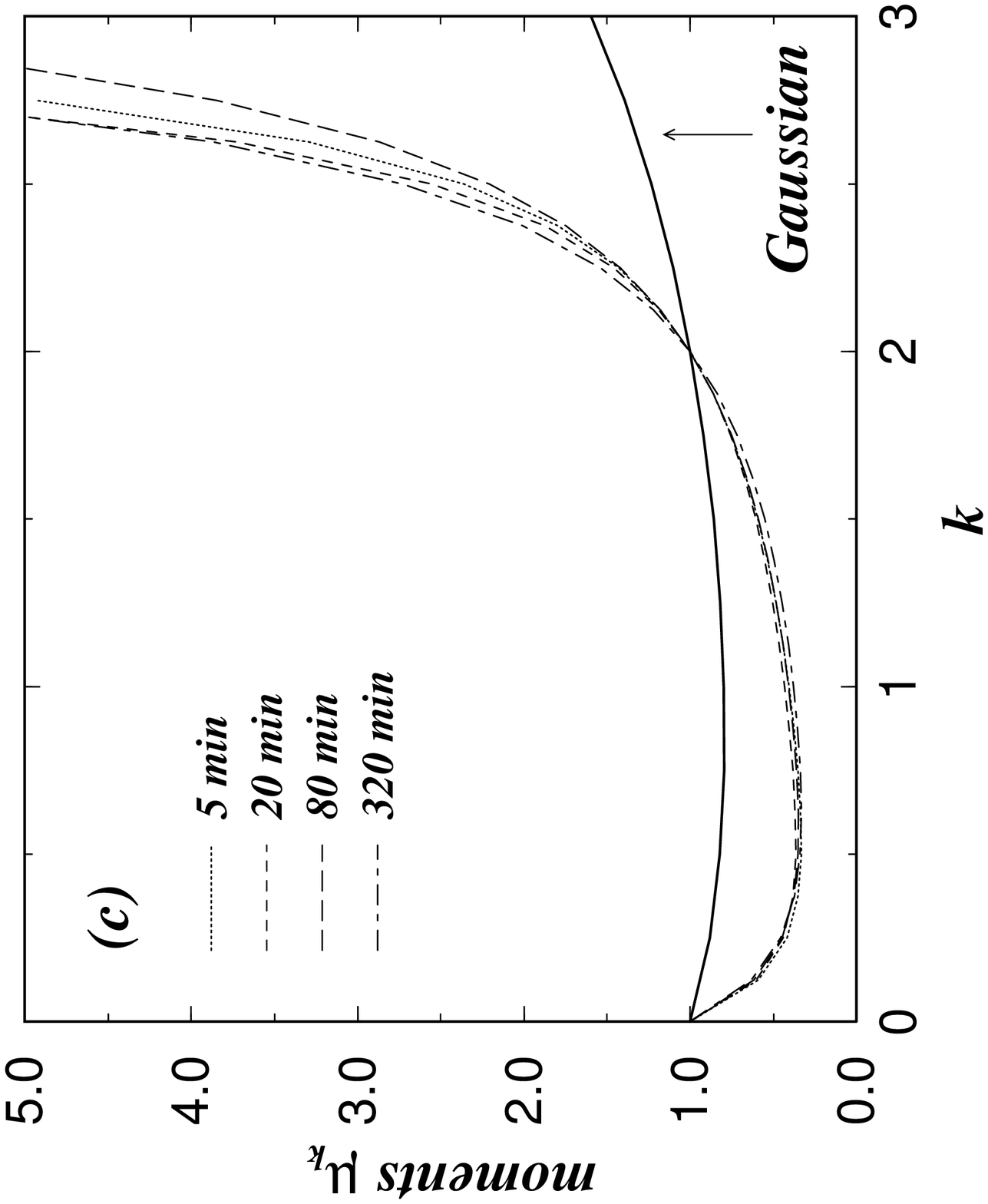}}}
}
\caption{{\it (a)} Cumulative distribution of normalized returns for
$\Delta t = 30$~min. The filled squares show the distribution for
returns normalized by the time-averaged volatility for each company,
as defined in Eq.~(\protect{\ref{eq_defg}}). The circles show the
distribution for returns normalized by the average volatility for each
size bin, Eq.~(\protect{\ref{eq_g_tcond}}), showing the consistency of
these two methods. {\it (b)} The distribution of returns for different
time scales $\Delta t \leq 1\,$day. The exponents from the power-law
regression fits are summarized in Table~\protect\ref{alpha.exp}. {\it
(c)} Fractional moments from $0\leq k < 3$ for the normalized returns
for the same scales as in {\it (b)}. Note that the moments are not
converging to Gaussian behavior, for example, at large $k$ the moments
for $\Delta t=80$~min is to the right of $\Delta t=320$~min. The thick
full line shows the Gaussian moments.}
\label{chk_taq}
\end{figure}

Figure~\ref{chk_taq}(a) shows the distribution of normalized 30~min
returns.  We test if our alternative procedure of normalizing the
returns by the time averaged volatility for each bin of market
capitalization $S$ is consistent with the previous procedure of
normalizing by the time averaged volatility for each company through
Eq.~(\ref{eq_defg}). To this end, we also show in
Fig.~\ref{chk_taq}(a) the distribution of normalized 30~min returns
using the normalization of Eq.~(\ref{eq_defg}). The distribution of
returns obtained by both procedures are consistent with a power law
decay of the same form as Eq.~(\ref{def_alpha}). Power-law regression
fits to the positive tail yield estimates of $\alpha= 3.21 \pm 0.08$
for the former method and $\alpha = 3.23 \pm 0.05$ for the latter,
confirming the consistency of the two procedures.  The values of the
exponent for 30~min time scales, $\alpha = 3.21 \pm 0.08$ (positive
tail) and $\alpha = 3.01 \pm 0.12$ (negative tail), are also
consistent with the estimates, Eq.~(\ref{alpha_val}), for 5 min
normalized returns.

Next, we compute the distribution of returns for longer time scales
$\Delta t$. Figure~\ref{chk_taq}(b) shows the cumulative distribution
of the normalized returns for time scales from 5~min up to 1~day. We
observe good ``data collapse'' with consistent values of $\alpha$
which suggests that the distribution of returns appears to retain its
functional form for larger $\Delta t$. The scaling of the distribution
of returns for individual companies is consistent with previous
results for the distribution of the S\&P 500 index
returns~\cite{Mantegna95,Gopi99}. The estimates of the exponent
$\alpha$ from power-law regression fits to the cumulative distribution
and from the Hill estimator are listed in
Table~\protect\ref{alpha.exp}.

\subsection{Scaling of the moments for $\Delta t < 1$~day}
%%%%%%%%%%%%%%%%%%%%%%%%%%%%%%%%%%%%%%%%%%%%%%%%

In the preceding subsection we reported that the distribution of
returns retains the same functional form for 5~min$< \Delta t<$
1~day. We can further test this scaling behavior by analyzing the
moments of the distribution of normalized returns $g$,
\begin{equation}
\mu_k \equiv \langle\, \vert g \vert^k \, \rangle\,,
\label{moments}
\end{equation}
where $\langle \dots \rangle$ denotes an average over all the normalized
returns for all the bins. Since $\alpha \approx 3$, we expect $\mu_k$ to
diverge for $k \geq 3$, and hence we compute $\mu_k$ for $k< 3$.

Figure~\ref{chk_taq}(c) shows the moments of the normalized returns
$g$ for different time scales from 5~min up to 1~day. The moments do
not vary significantly for the above time scales, thus confirming the
scaling behavior of the distribution observed in Fig~\ref{chk_taq}(b).

%%%%%%%%%%%%%%%%%%%%%%%%%%%%% SECTION
\section{The distribution of returns for 1~day~$\leq \Delta t \leq 16$~days}
%%%%%%%%%%%%%%%%%%%%%%%%%%%%%%%%%%%%%%%%%%%%%%%%
 
For time scales of 1$\,$day or longer, we analyze data from the CRSP
database. We analyze approximately $3.5 \times 10^7$ daily records for
about 16,000 companies for the 35-year period 1962-96. We expect the
market capitalization of a company to change dramatically in such a
long period of time. Further, we expect small companies to be more
volatile than large companies. Hence, large changes that might occur
in the market capitalization of a company will lead to large changes
on its average volatility.  To control for these changes in market
capitalization, we adopt the method that was used in the previous
subsection for $\Delta t >$ 5~min.

Thus, we compute the cumulative conditional probability $P(G_S \geq x |
S)$ that the   return $G_S \equiv G_S(t,\Delta t)$ is greater than
$x$, for a given bin of average market capitalization $S$.  We first
divide the entire range of $S$ into bins of uniform length in
logarithmic scale. We then compute a separate probability distribution
for the returns $G_S$ which belong to a bin of average market
capitalization $S$.

\begin{figure}
\narrowtext
\centerline{
\epsfysize=0.5\columnwidth{\rotate[r]{\epsfbox{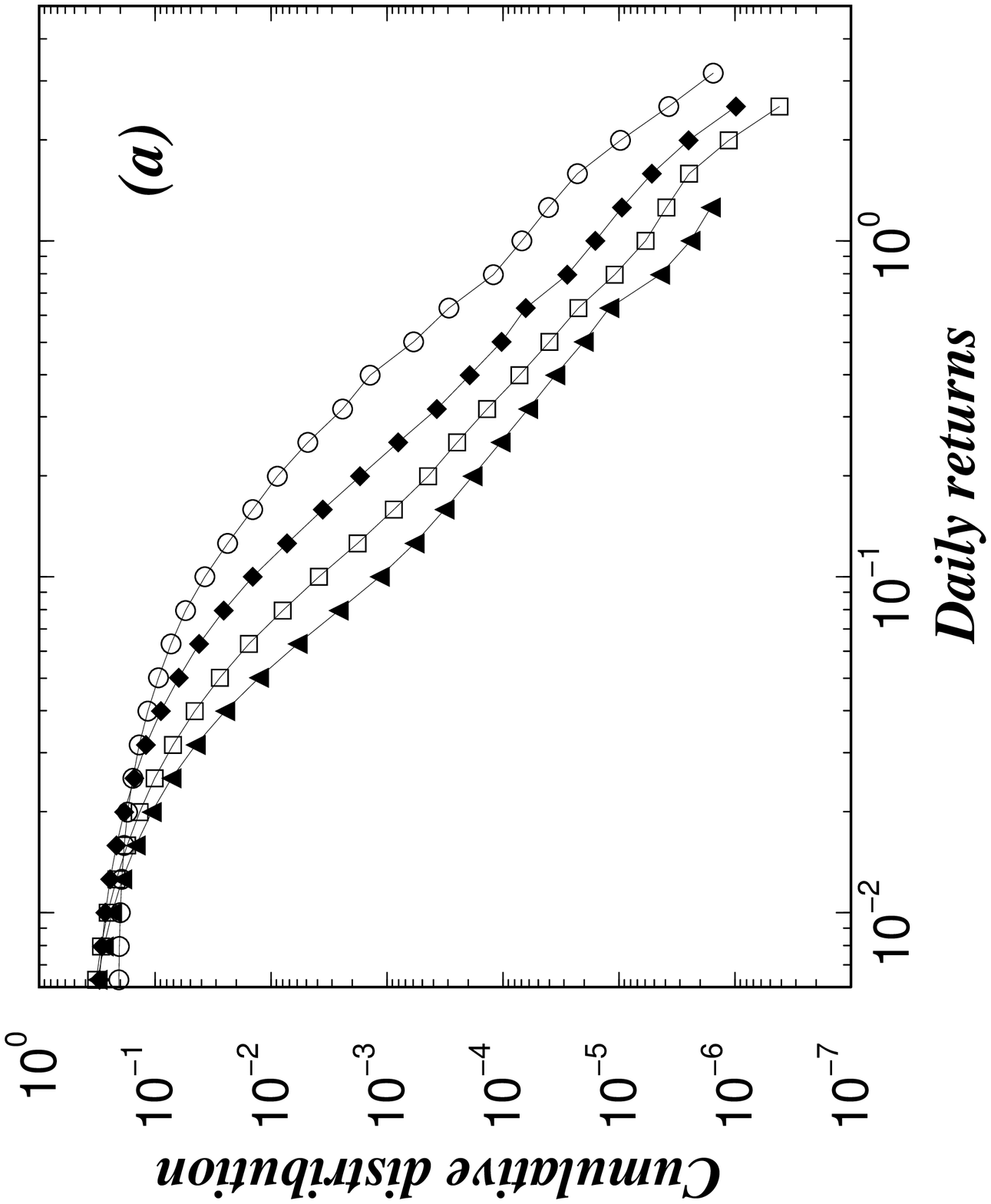}}}
\hspace*{0.5cm}
\epsfysize=0.5\columnwidth{\rotate[r]{\epsfbox{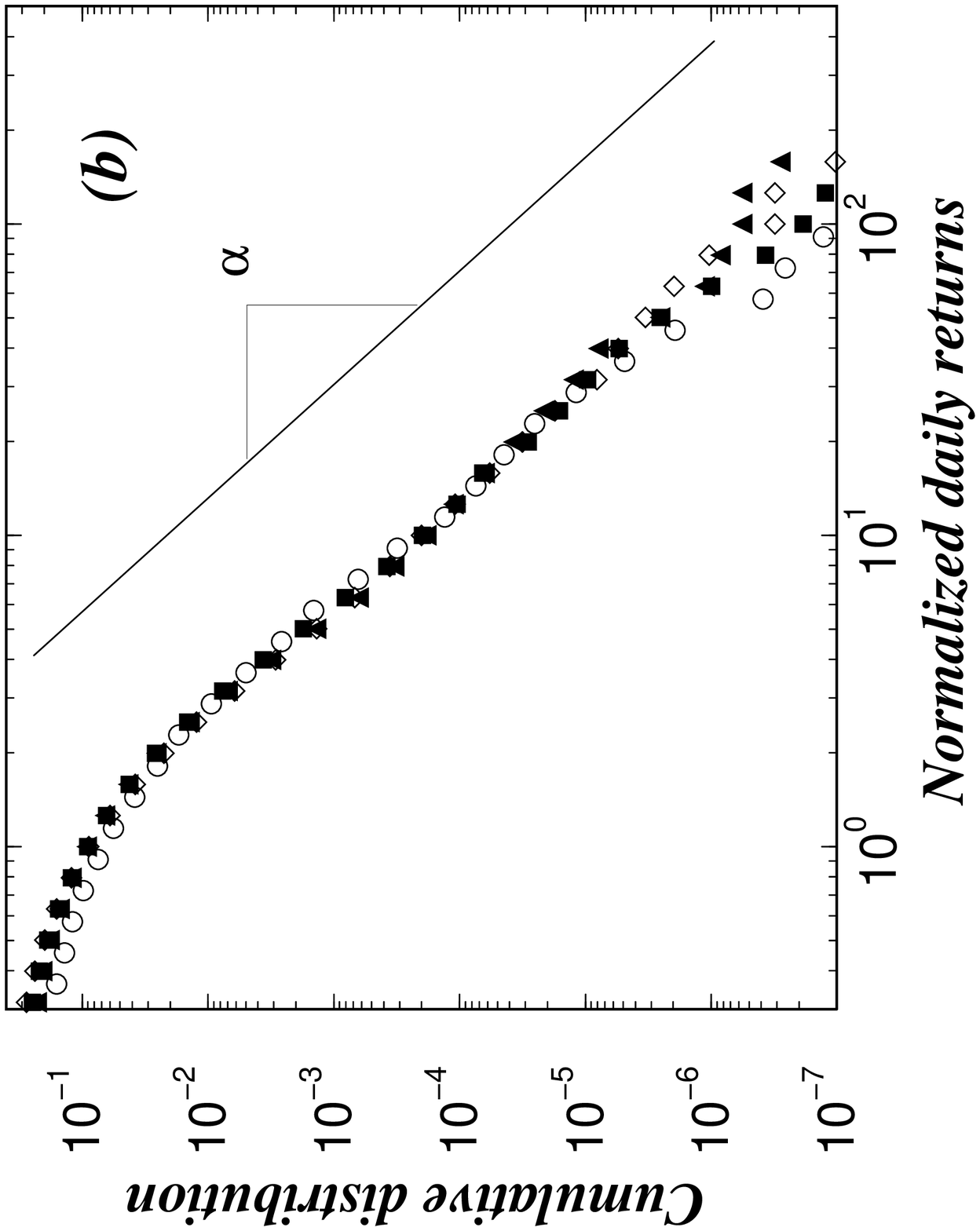}}}
}
\vspace{0.5cm}
\caption{ {\it (a)} Cumulative distribution of the conditional
probability $P(g > x | S)$ of the returns for companies with starting
values of market capitalization S for $\Delta t$ = 1$\,$day from the
CRSP database.  We define uniformly spaced bins on a logarithmic
scale and  show the distribution of returns for the bins, $10^5 < S
\leq 10^6$, $10^6 < S \leq 10^7$, $10^7 < S \leq 10^8$, and $10^8 < S
\leq 10^9$. {\it (b)} Cumulative conditional distributions of
returns normalized by the average volatility $v_S (\Delta t)$ of each
bin.}
\label{indcrsp}
\end{figure}

Figure~\ref{indcrsp}(a) shows the cumulative distribution of daily
returns $P(G_S > x | S)$ for different values of $S$. Since the widths
of these distributions are different for different $S$, we analyze the
normalized returns $g_S$, which were defined in Eq.~(\ref{eq_g_tcond}).

Figure~\ref{indcrsp}(b) shows the cumulative distribution $P(g_S>x)$
of the normalized daily returns $g_S$. These distributions appear to
have similar functional forms for different values of $S$. In order to
improve statistics, we compute a {\it single} cumulative distribution
$P(g_S>x)$ of the normalized returns for all $S$. We observe a
power-law behavior of the same form as Eq.~(\ref{def_alpha}).
Regression fits yield estimates for the exponent, $\alpha =2.96 \pm
0.09$ for the positive tail and $\alpha =2.70\pm0.10$ for the negative
tail.

\begin{figure}
\narrowtext
\centerline{
\epsfysize=0.5\columnwidth{\rotate[r]{\epsfbox{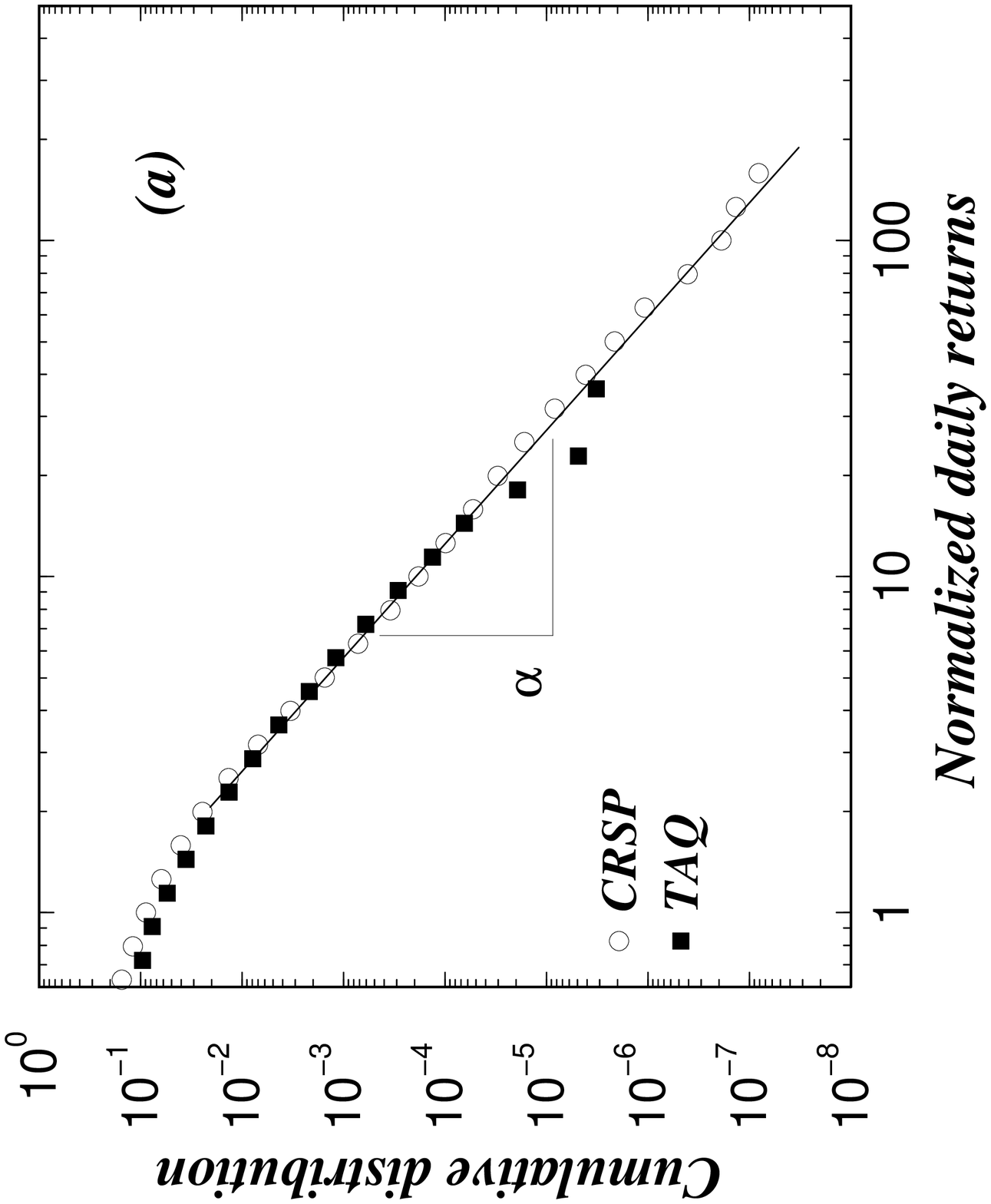}}}
\hspace{0.5cm}
\epsfysize=0.5\columnwidth{\rotate[r]{\epsfbox{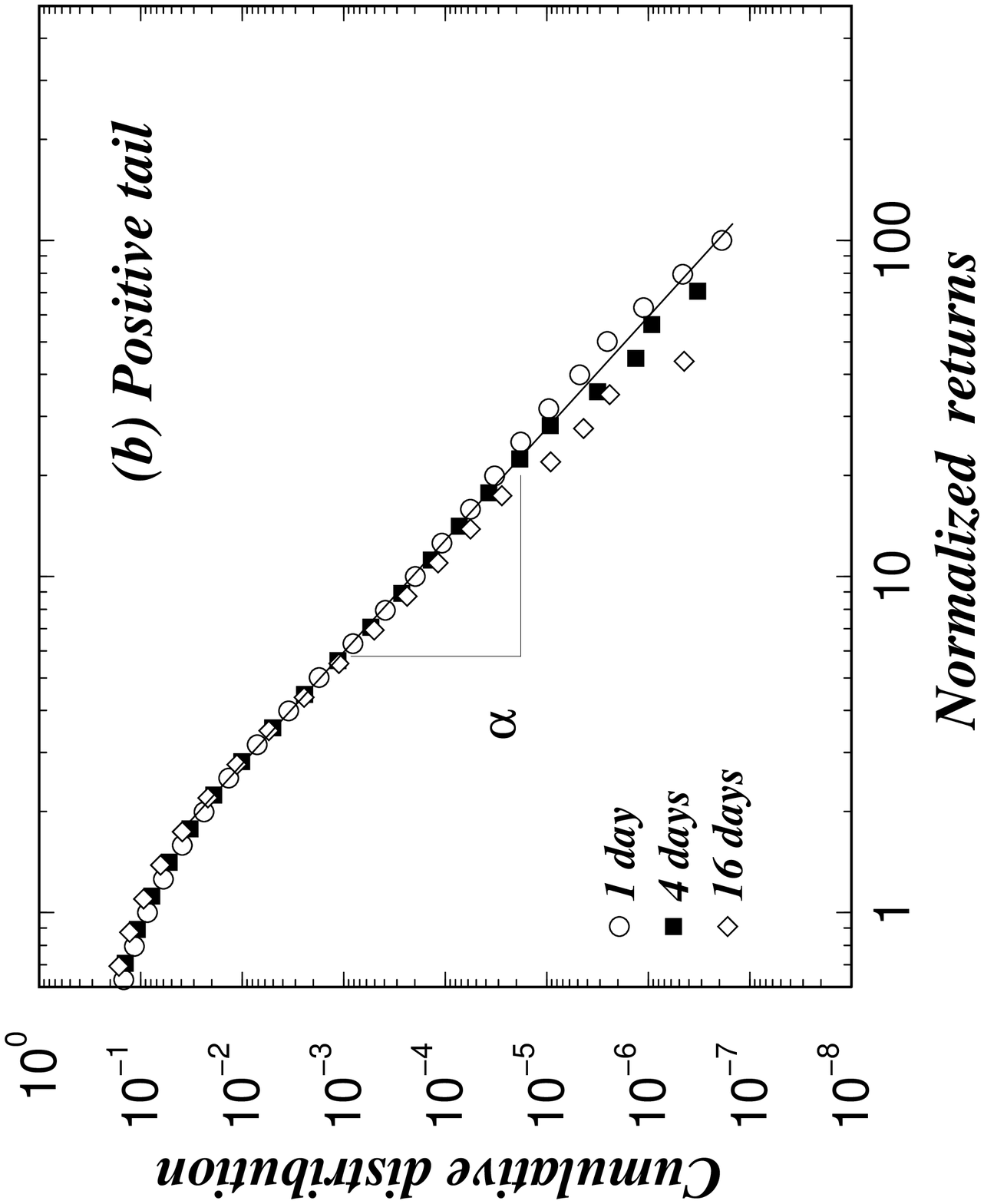}}} 
}
\centerline{
\epsfysize=0.5\columnwidth{\rotate[r]{\epsfbox{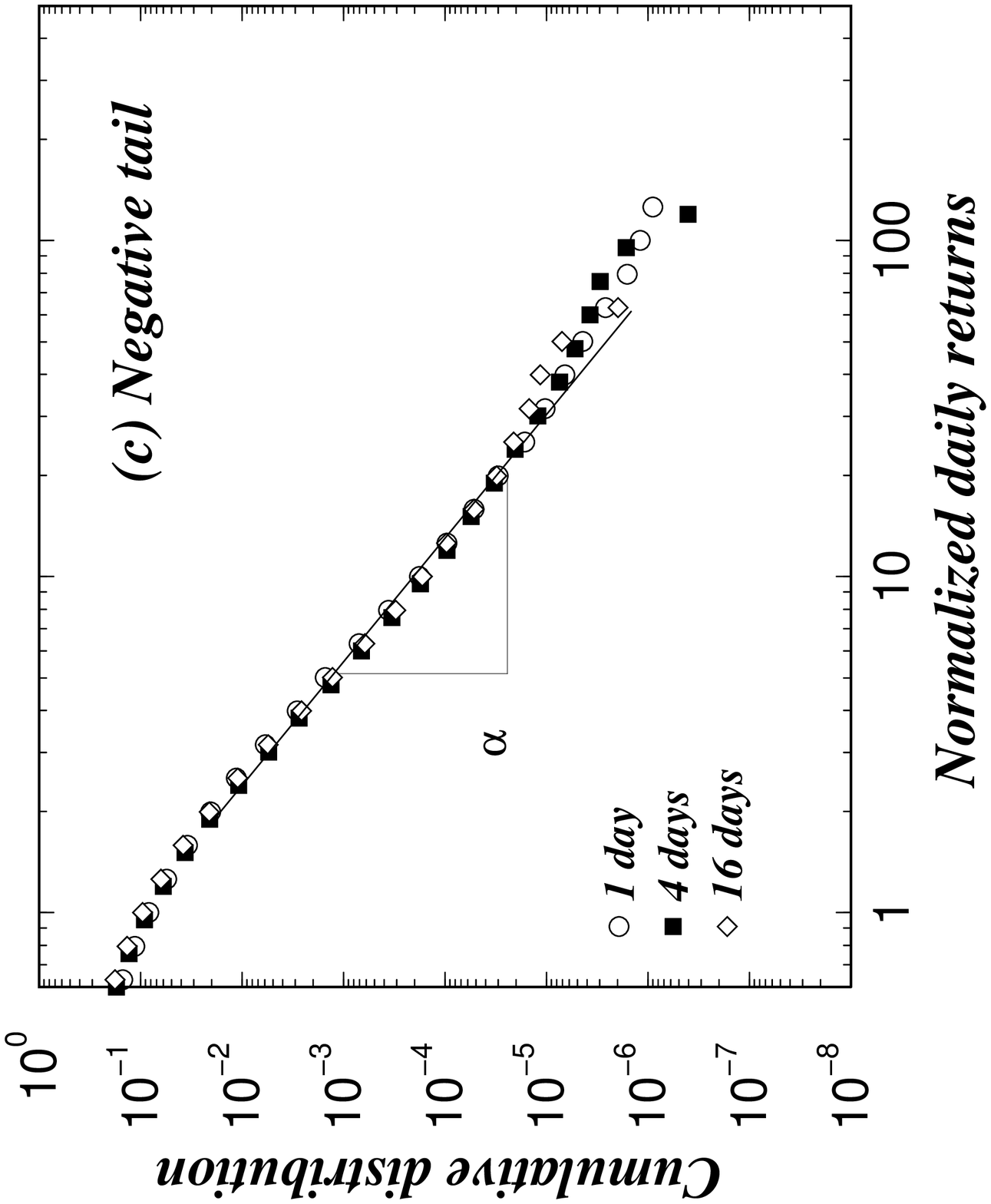}}}
\hspace{0.5cm}  
\epsfysize=0.5\columnwidth{\rotate[r]{\epsfbox{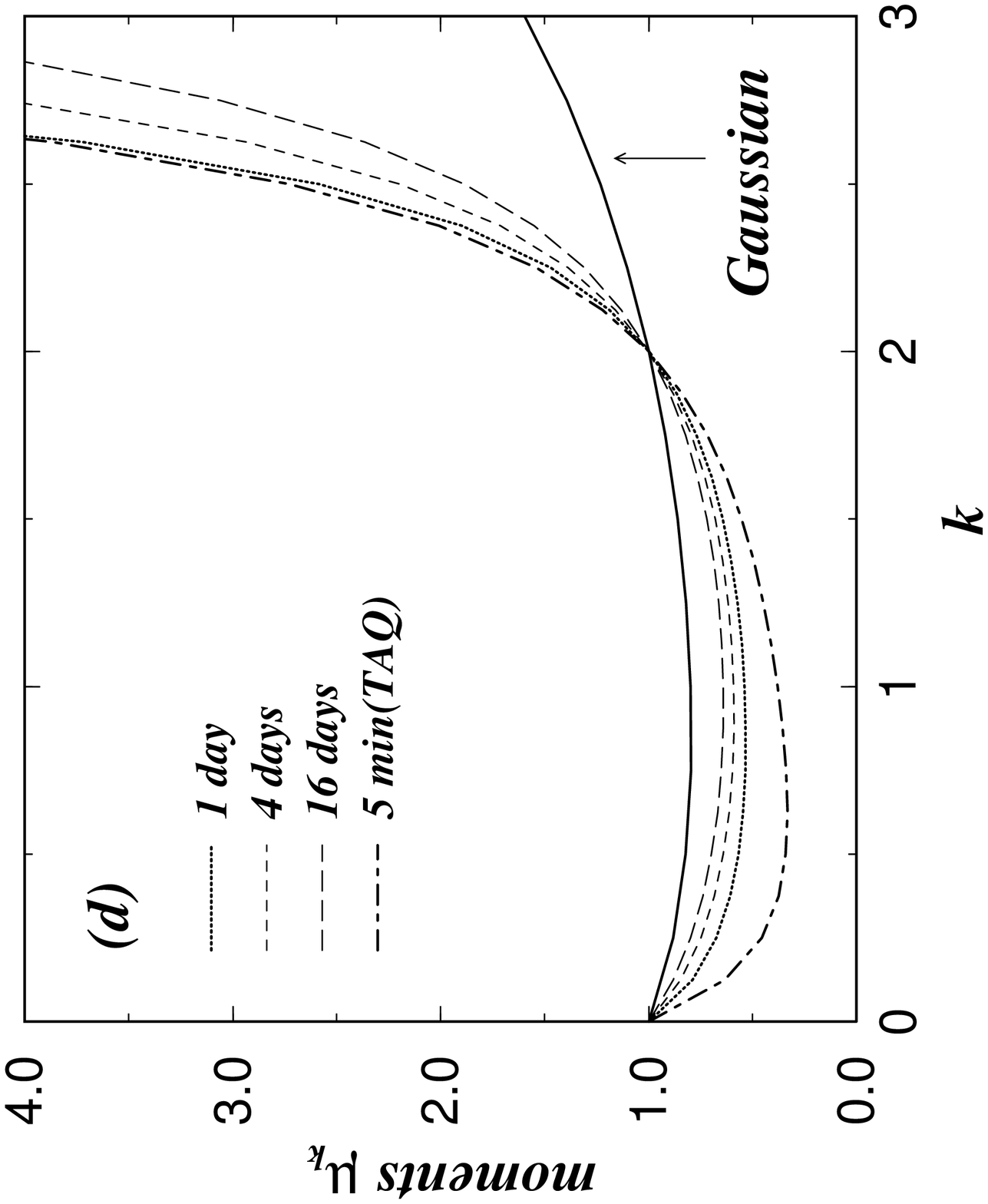}}}
}
\vspace{0.2cm}
\caption{{\it (a)} Cumulative distribution of normalized daily returns
computed from the CRSP database contrasted with the same distribution
from the TAQ database, normalized by the average
volatility. Regression fits yield estimates $\alpha =2.96 \pm 0.09$
(positive tail), and $\alpha =2.70 \pm 0.10$ (negative tail) for the
CRSP data, and $\alpha=3.27 \pm 0.19 $ (positive tail) and
$\alpha=2.98 \pm 0.21$ (negative tail) for the TAQ data. The
regression fits were performed for the region $2 \leq g \leq 80$. {\it
(b)} Positive and {\it (c)} negative tails of the cumulative
distribution of normalized returns for $\Delta t = 1, 4$ and 16~days.
Estimates of the exponents are listed in
Table~\protect\ref{alpha.exp}.  {\it (d)} The fractional moments
$\mu_k \equiv \langle \vert g \vert^k \rangle$ for the normalized
returns for the same time scales.The thick full line shows the
Gaussian moments.
%where $\langle \dots \rangle$ denotes an average over all the normalized
%returns. 
}
\label{allcrsp}
\end{figure}

Figure~\ref{allcrsp}(a) compares the cumulative distributions of the
normalized 1~day returns obtained from the CRSP and TAQ databases. The
estimates of the power-law exponents obtained from regression fits are
in good agreement for these two databases.

Figures~\ref{allcrsp}(b,c) show the distributions of normalized
returns for $\Delta t = 1, 4,16$~days. The estimates of the exponent
$\alpha$ increase slightly in value for the positive tail, while for
the negative tail the estimates of $\alpha$ are approximately
constant. The increase in $\alpha$ for the positive tail is also
reflected in the moments [Fig.~\ref{allcrsp}(d)].

%%%%%%%%%%%%%%%%%%%%%%%%%%%%% SECTION
\section{The distribution of returns for $\Delta t \geq 16$~days}
%%%%%%%%%%%%%%%%%%%%%%%%%%%%%%%%%%%%%%%%%%%%%%%%

\begin{figure}
\narrowtext
\centerline{
\epsfysize=0.5\columnwidth{\rotate[r]{\epsfbox{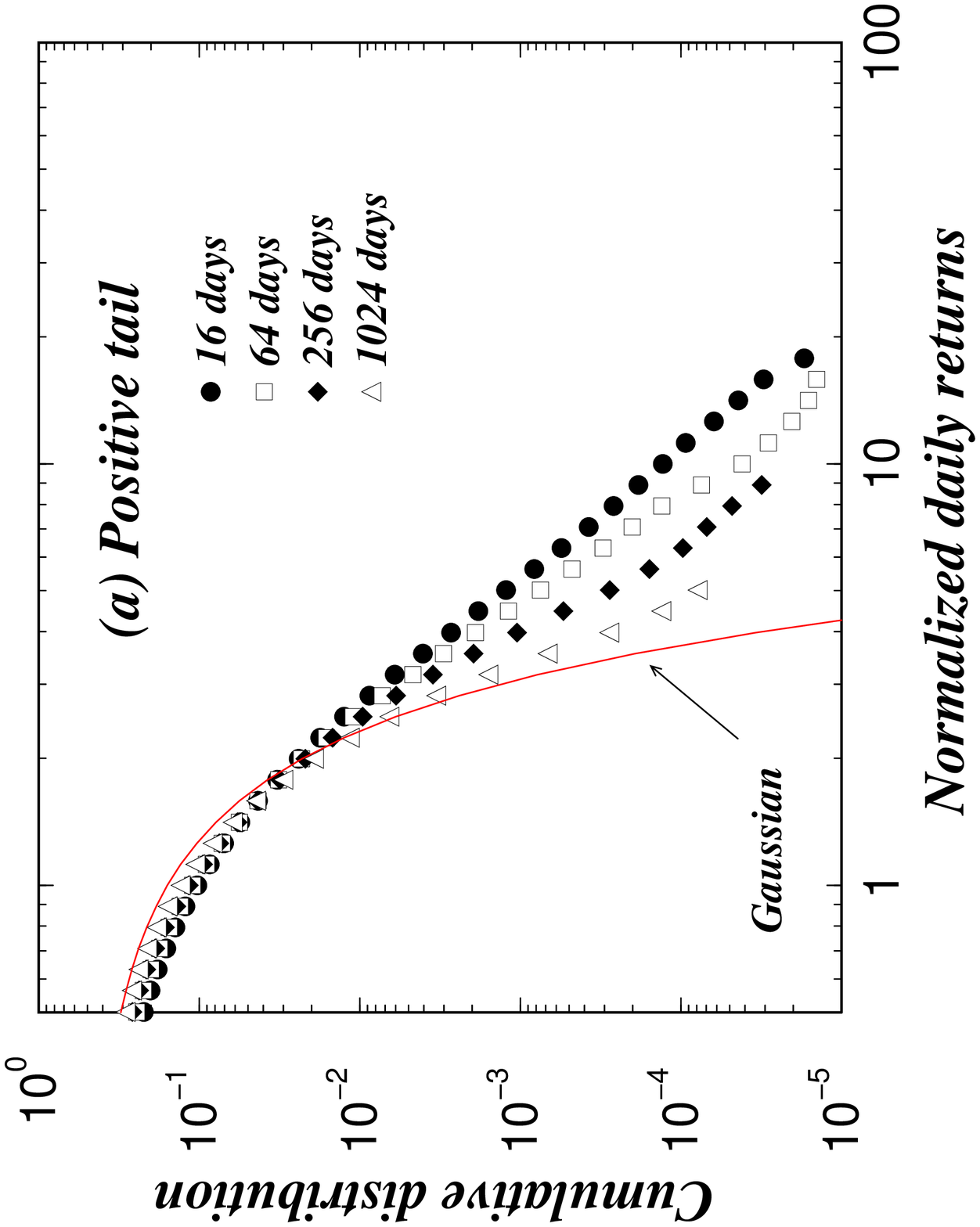}}}
\hspace{0.5cm}
\epsfysize=0.5\columnwidth{\rotate[r]{\epsfbox{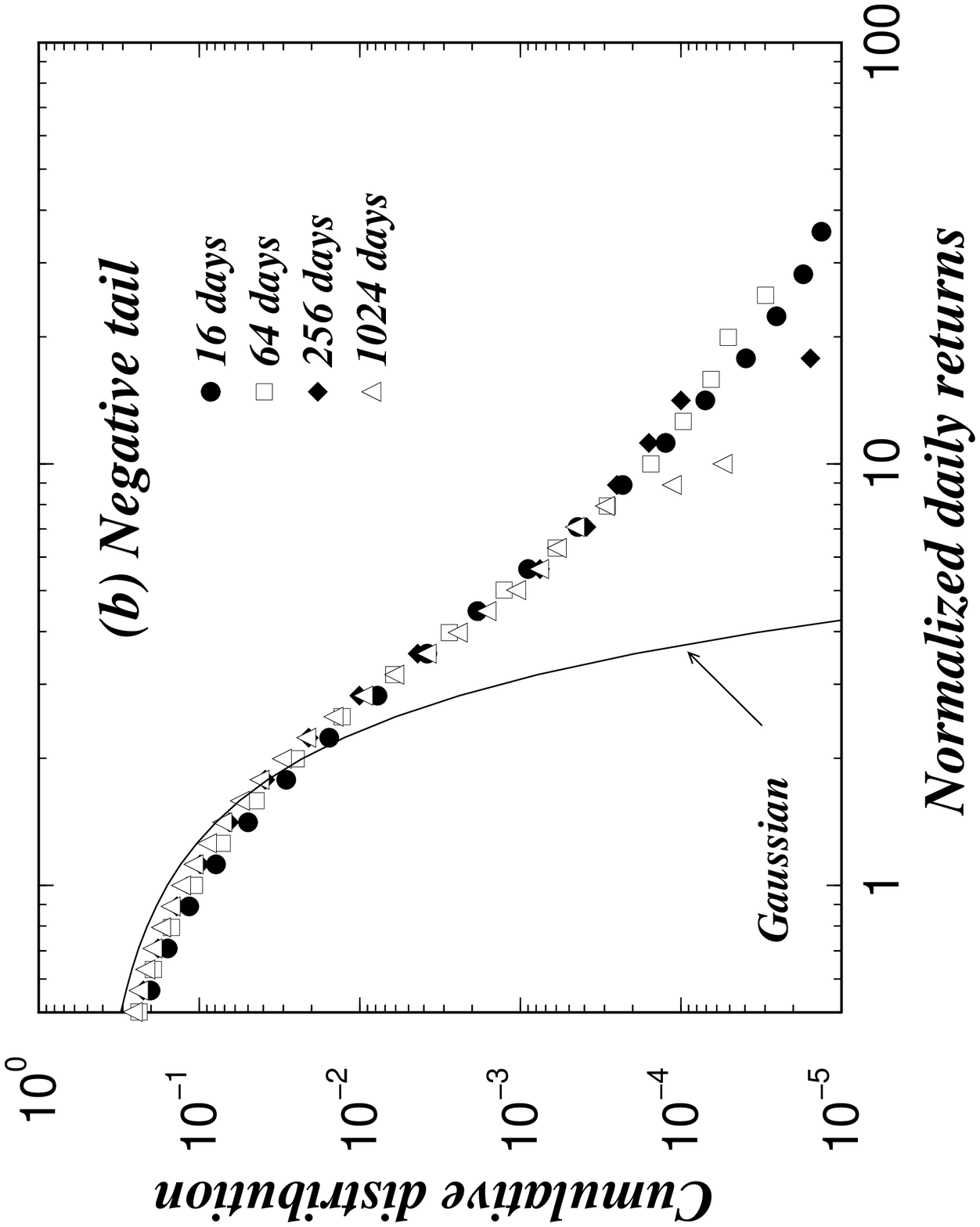}}}
}
\vspace{0.5cm}
\centerline{
\epsfysize=0.5\columnwidth{\rotate[r]{\epsfbox{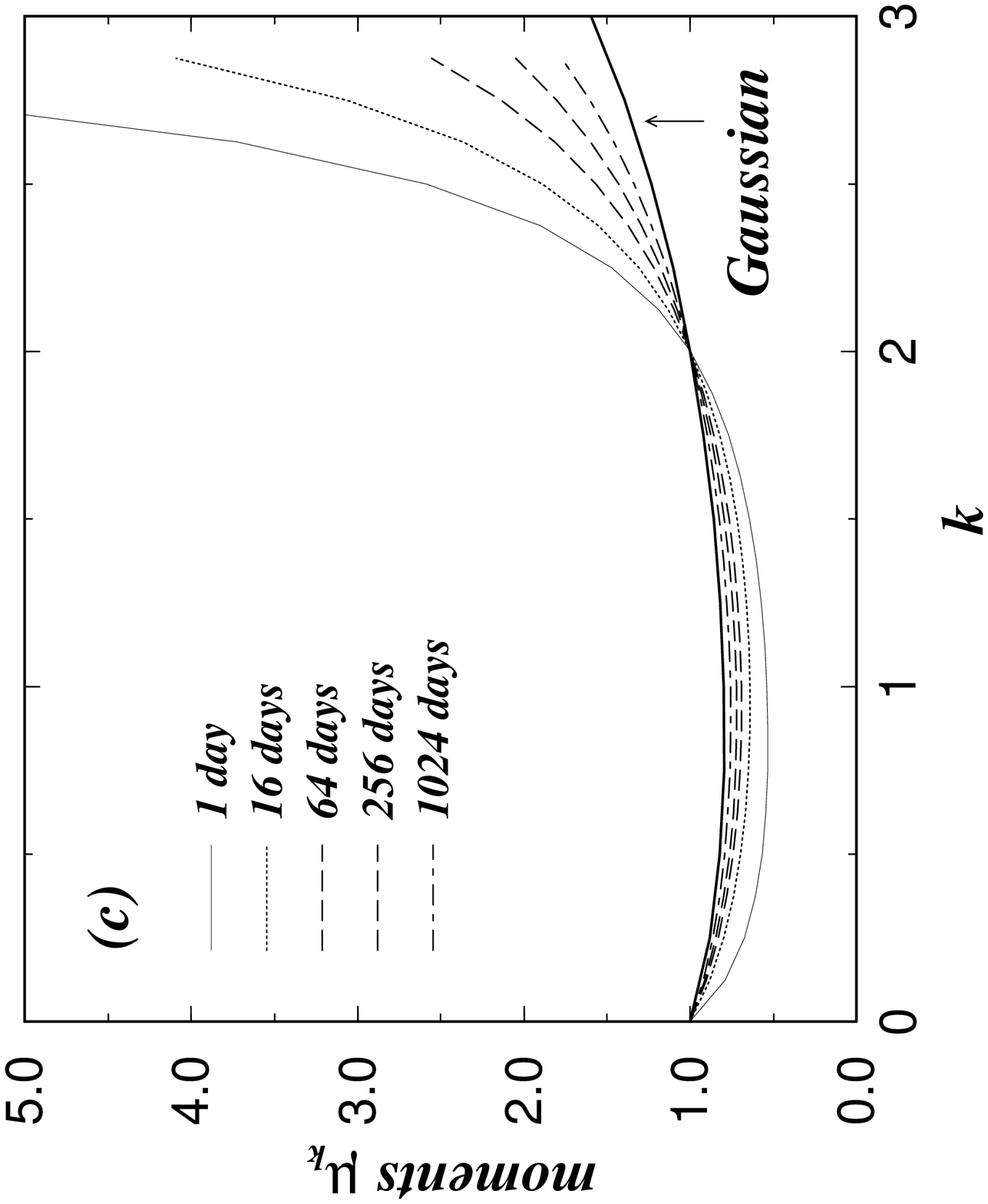}}}
}
\vspace*{0.5cm}
\caption{ {\it (a)} Positive and {\it (b)} negative tails of the
cumulative distribution of the normalized returns for $\Delta t = 16,
64, 256$ and 1024~days. The positive tail shows clear indication of
convergence to Gaussian behavior, whereas for the negative tail the
power-law behavior still seems to hold, although the statistics at the
tail are limited for the longer time scales. 
Estimates of the exponents are listed in Table~\protect\ref{alpha.exp}.
{\it (c)} The fractional moments $\mu_k,\, 0 \leq k < 3 $ of the normalized
returns for $\Delta t = 16, 64, 256$ and 1024~days show clear indication of
convergence to Gaussian behavior with increasing $\Delta t$. }
\label{scl_brk}
\end{figure}

The scaling behavior of the distributions of returns appears to break
down for $\Delta t \geq 16$~days, and we observe indications of slow
convergence to Gaussian behavior.  In Figs.~\ref{scl_brk}(a,b) we show
the cumulative distributions of the normalized returns for $\Delta t
\geq 16$~days. For the positive tail, we find indications of
convergence to a Gaussian, while the negative tail appears not to
converge. The convergence to Gaussian behavior is also apparent
from the behavior of the moments for these time scales
[Fig.~\ref{scl_brk}(c)].

\begin{figure}
\narrowtext 
\centerline{
\epsfysize=0.5\columnwidth{\rotate[r]{\epsfbox{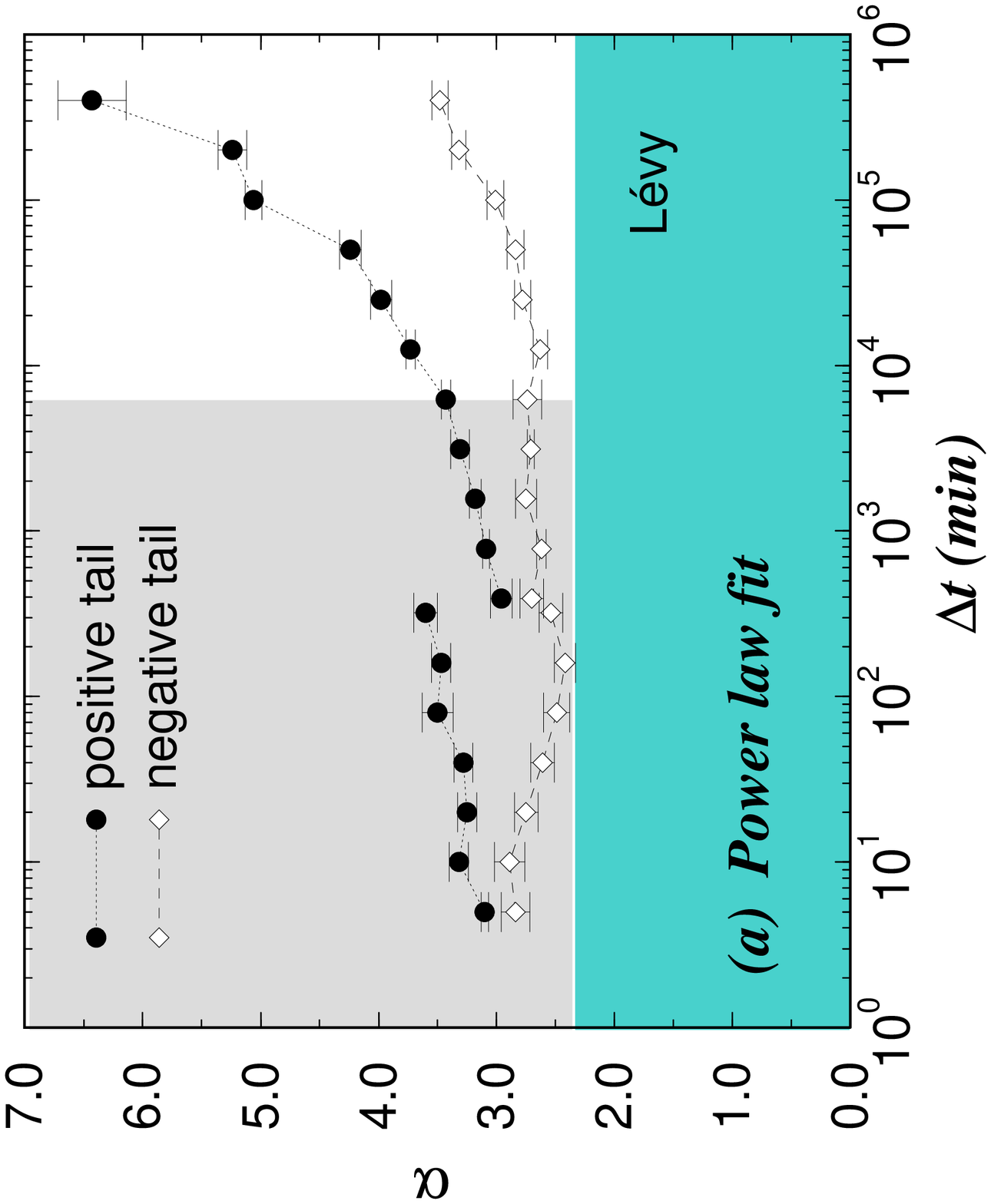}}}
\hspace{0.5cm}
\epsfysize=0.48\columnwidth{\rotate[r]{\epsfbox{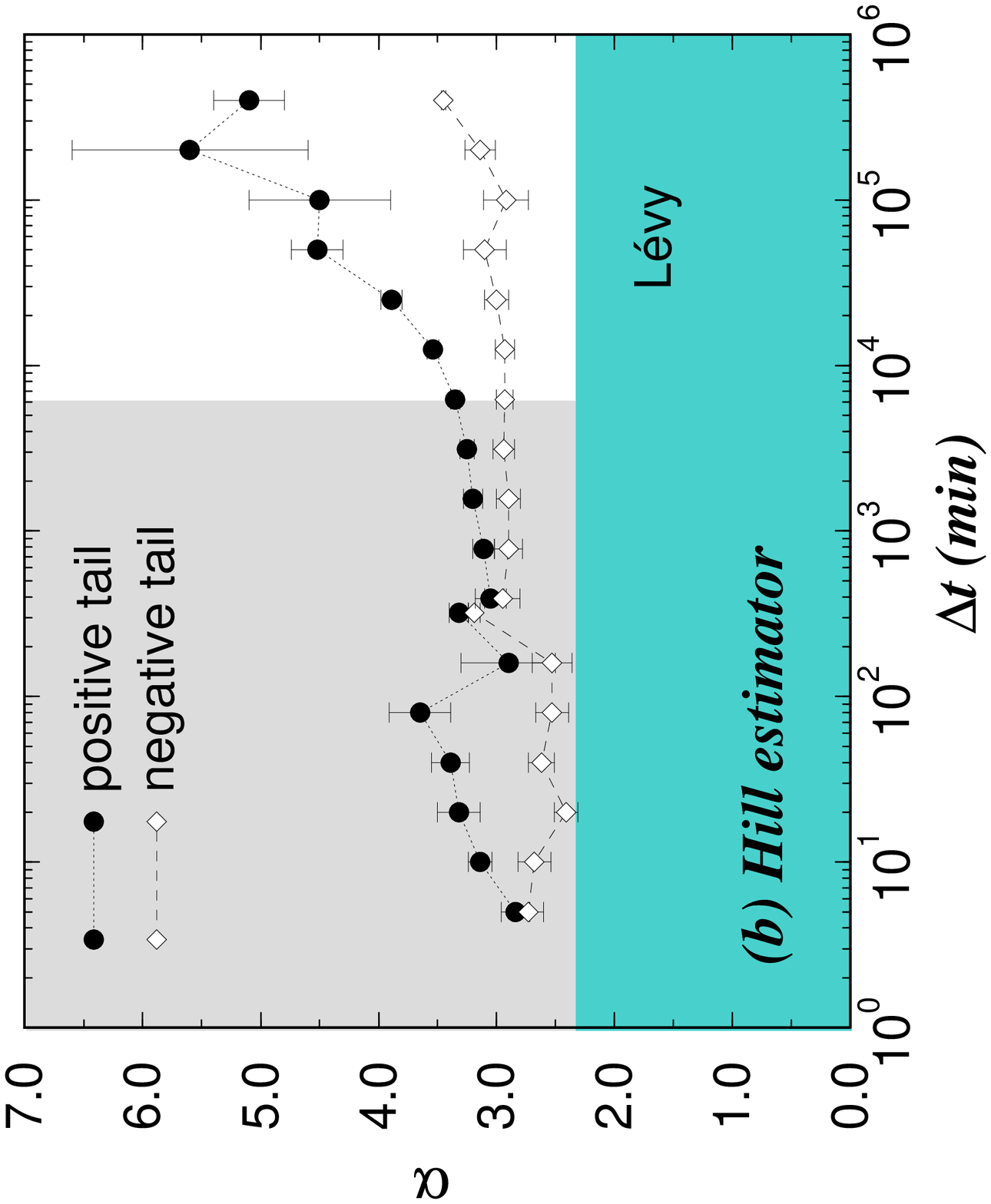}}}
}
\caption{ The values of the exponent $\alpha$ characterizing the
asymptotic power-law behavior of the distribution of returns as a
function of the time scale $\Delta t$ obtained using (a) a power-law
fit, and (b) the Hill estimator. The values of $\alpha$ for $\Delta t
< $1~day are calculated from the TAQ database while for $\Delta t \geq
$1~day they are calculated from the CRSP database. The unshaded
region, corresponding to time scales larger than $(\Delta t)_{\times}
\approx 16$~days (6240~min), indicates the range of time scales where
we find results consistent with slow convergence to Gaussian behavior.}
\label{alpha.fig}
\end{figure}

To summarize our results for the distribution of individual company
returns, we find that (i) the distribution of normalized returns for
individual companies is consistent with a power-law behavior
characterized by an exponent $\alpha \approx 3$, (ii) the
distributions of returns retain the same functional form for a wide
range of time scales $\Delta t$, varying over 3 orders of magnitude,
5~min$\leq \Delta t \leq 6240$~min = 16~days, and (iii) for $\Delta t
> 16$~days, the distribution of returns appears to slowly converge to
a Gaussian [Fig.~\ref{alpha.fig}].

%%%%%%%%%%%%%%%%%%%%%%%%%%%%% SECTION
\section{Cross-Correlations}
%%%%%%%%%%%%%%%%%%%%%%%%%%%%%%%%%%%%%%%%%%%%%%%%

In this section we address the second question that we posed
initially. That is, why is it that the distribution of returns for
individual companies and for the S\&P 500 index have the same asymptotic
form?  In the previous sections, we presented evidence that the
distribution of returns scales for a wide range of time intervals.  In
a previous study~\cite{Gopi99}, we demonstrated that this scaling
behavior is possibly due to time dependencies, in particular,
volatility correlations. Next, we will show that as the time
correlations lead to the time scaling of the distributions of returns,
so do cross correlations among different companies lead to a
functional form of the distribution of returns of indices similar to
that for single companies.

A direct way of analyzing the cross-correlations is by computing the
cross-correlation matrix~\cite{rmbs,laloux99,plerou}. Here, we take a
different approach, by analyzing the distribution of returns as a
function of market capitalization.

\begin{figure}
\narrowtext
\centerline{
\epsfysize=0.5\columnwidth{\rotate[r]{\epsfbox{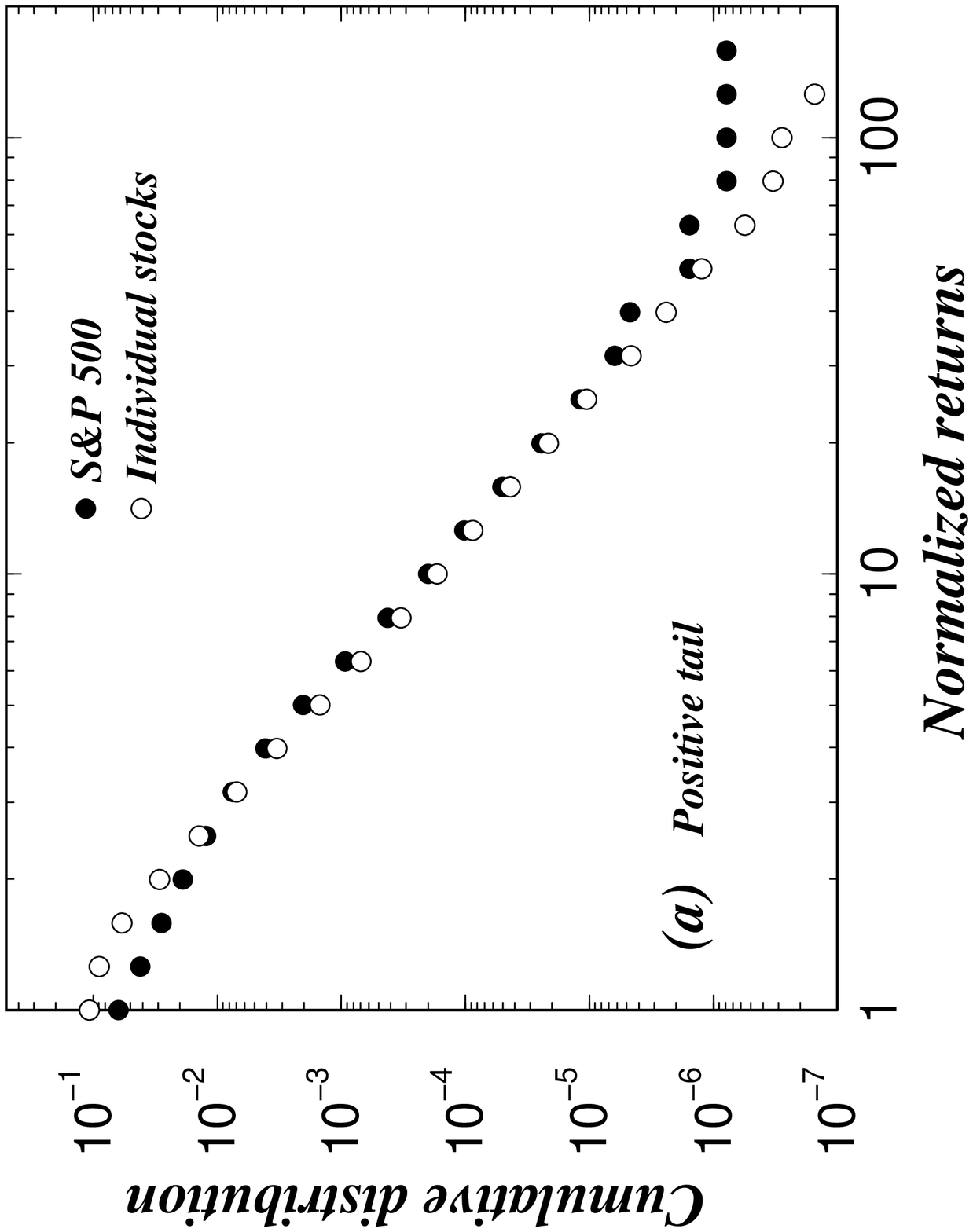}}}
\hspace{0.5cm}
\epsfysize=0.5\columnwidth{\rotate[r]{\epsfbox{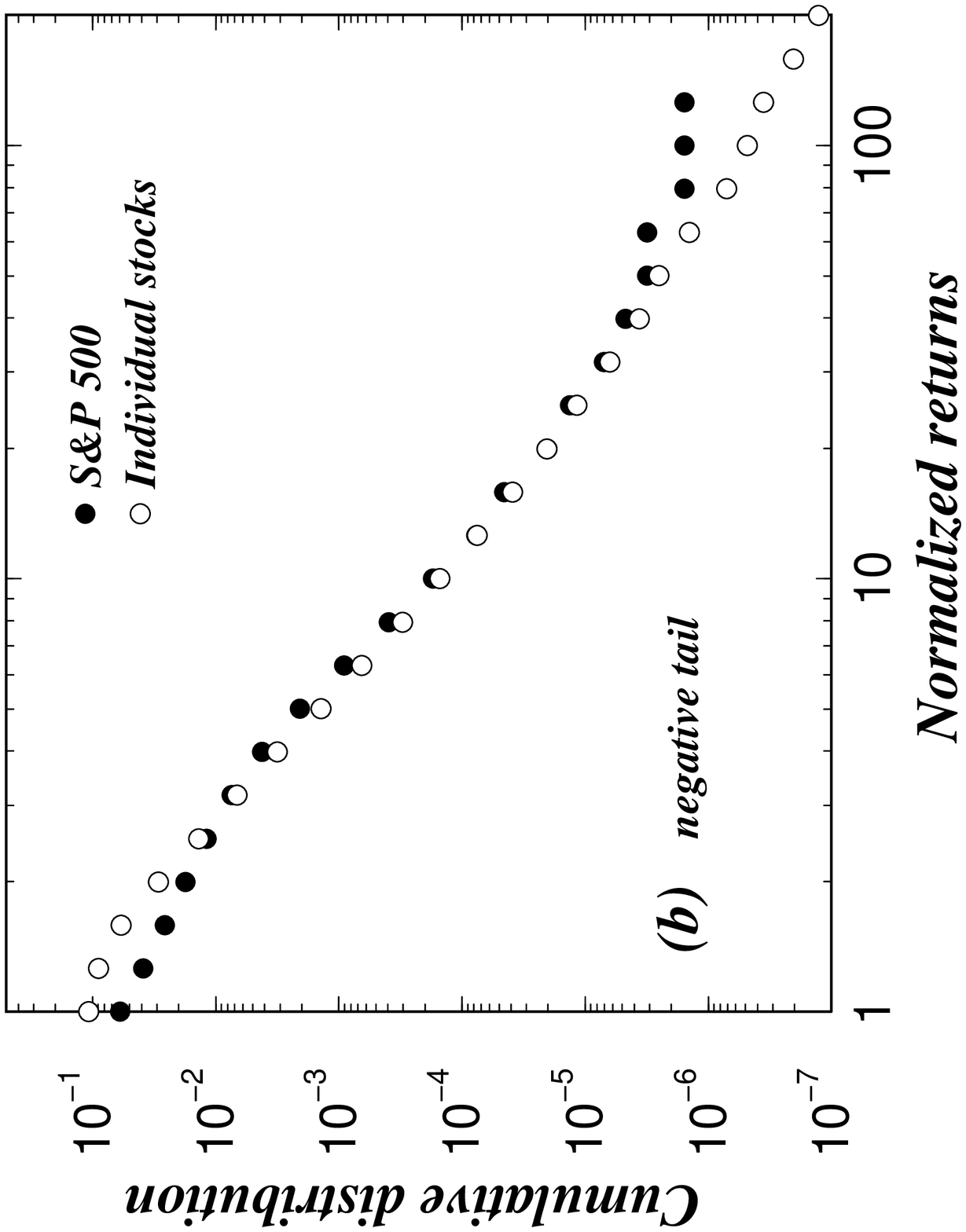}}} 
}
\vspace{0.5cm}
\centerline{
\epsfysize=0.5\columnwidth{\rotate[r]{\epsfbox{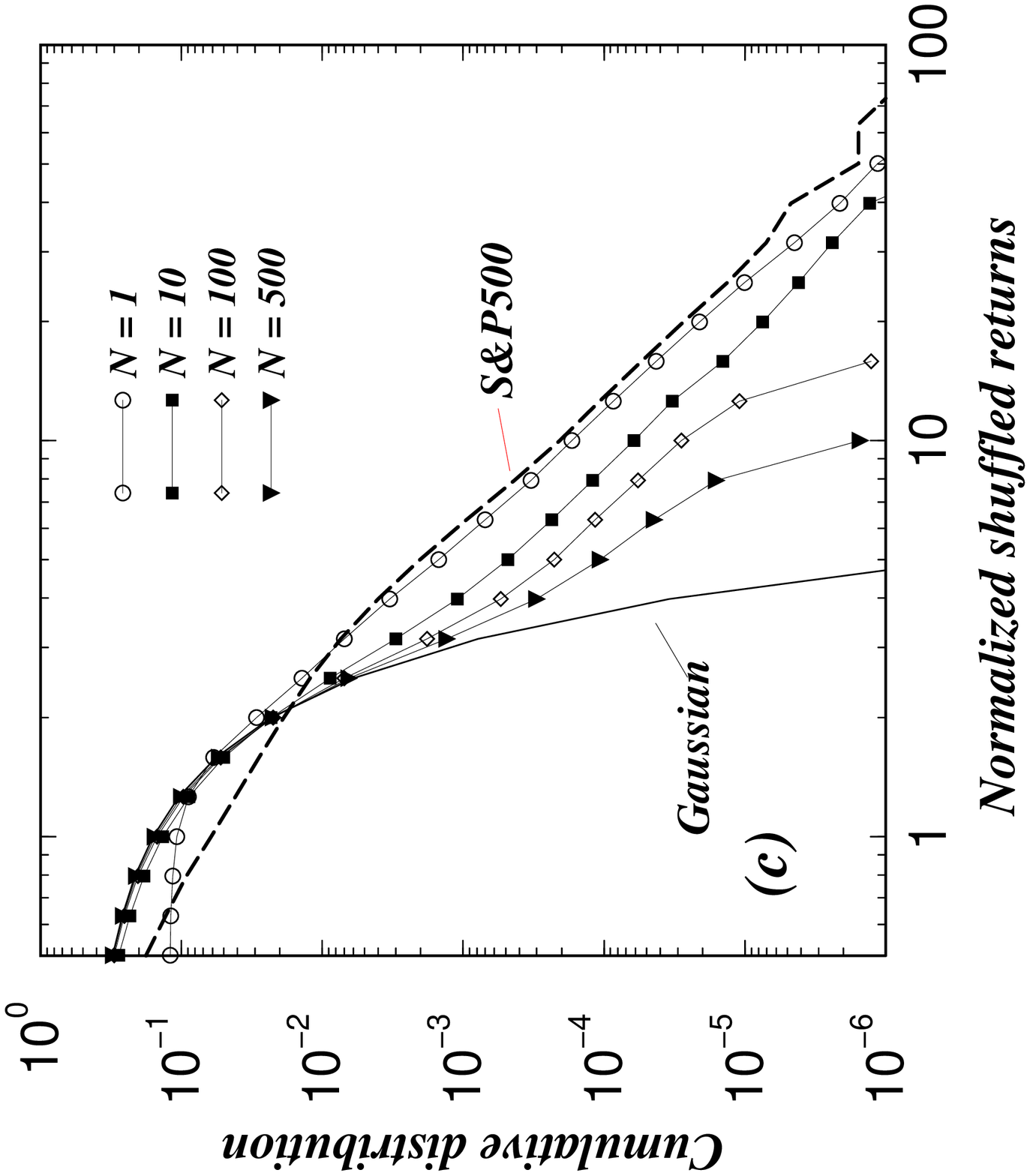}}}
}
\caption{{\it(a)} Positive and {\it (b)} negative tails of the
cumulative distribution for the normalized returns for the individual
companies and the S\&P 500 index. Both the distributions show the same
functional form, in spite of being a non-stable law. {\it(c)}
Cumulative distribution for the shuffled returns $\tilde
g^{(N)}(t,\Delta t)$ for $N=1,10,100,500$. The dotted curve is the
cumulative distribution for the S\&P 500. With increasing $N$ the
curves progressively approach a Gaussian, implying that without the
cross-dependencies between companies, the cumulative distribution for
the S\&P 500 would be almost Gaussian. }
\label{xcorr}
\end{figure}

First, we compare the distributions of the S\&P 500 index and that of
individual companies.  Figures~\ref{xcorr}(a,b) show the cumulative
distribution $P(g \geq x) $ for individual companies and for the S\&P
500 index.  The distributions show the same power-law behavior for $2
\le g \le 80$. This is surprising, because the distribution of index
returns $G_{SP500}(t,\Delta t)$ does not show convergence to Gaussian
behavior---even though the 500 distributions of individual returns
$G_i(t,\Delta t)$ are not stable.  Consider the family of index
returns defined as the partial sum \cite{sp_diff}
\begin{equation}
G_{(N)}(t,\Delta t) \equiv \sum_{i=1}^N w_i \, G_i(t,\Delta t)\,,
\end{equation}
where the weights $w_i\equiv S_i/\sum_{j=1}^N S_j$ have weak time
dependencies\cite{weights}. From the central limit theorem for random
variables with finite variance, we expect that the probability
distribution of $G_{(N)}$ would change systematically with $N$ and
approach a Gaussian for large $N$, provided there are no significant
dependencies among the returns $G_i$ for different $i$. Instead, we
find that the distribution of $G_{(N)}$ has the same asymptotic
behavior as that for individual companies.

In order to show that the scaling behavior may be due to
cross-correlations between companies, we first destroy any existing
dependencies among the returns of different companies by randomizing
each of the 1000 time series $G_i(t)$. By adding up the shuffled
series, we construct a shuffled index return $ G_{(N)}^{sh}(t)$ out of
statistically independent companies with the same distribution of
returns. Fig.~\ref{xcorr}(c) shows the cumulative distribution of the
shuffled index returns $G_{(N)}^{sh}(t,\Delta t)$ for increasing $N$
and $\Delta t = 5$~min. The distribution changes with $N$, and
approaches a Gaussian shape for large $N$, which indicates that the
scaling in Fig.~\ref{xcorr}(a) is caused by non-trivial dependencies
between different companies. 

%Furthermore, one possible explanation for the power-law dependence of
%the average fluctuation of returns as a function of the ``size'',
%measured by the market capitalization [Fig.~\ref{varsize}(a)], could
%be the existence of cross-correlations of similar structure as in
%scale-invariant systems. The structure of these correlations is
%reminiscent of critical phenomena, where the long range correlations
%between interacting units at the critical point result in scale
%invariant properties\cite{Stanley71}.

%%%%%%%%%%%%%%%%%%%%%%%%%%%%% SECTION
\section{Discussion}
%%%%%%%%%%%%%%%%%%%%%%%%%%%%%%%%%%%%%%%%%%%%%%%%

We have presented a systematic analysis, on two different databases,
of the distribution of returns for individual companies for time
scales $\Delta t$ ranging from 5$\,$min up to $\approx 4$~years.  We
find that the distribution of returns is consistent with a power-law
asymptotic behavior, characterized by an exponent $\alpha \approx
3$---well outside the stable L\'evy regime $0 < \alpha < 2$---for time
scales up to approximately 16~days. For longer time scales, the
scaling behavior appears to break down and we observe ``slow''
convergence to Gaussian behavior.

We also find that the distribution of returns of individual
companies and the S\&P 500 index have the same asymptotic
behavior. This scaling behavior does not hold when the
cross-correlations between companies are destroyed, suggesting the
existence of correlations between companies ---as occurs in strongly
interacting physical systems where power-law correlations at the
critical point result in scale-invariant properties. Recent studies of
the cross-correlation matrix using methods of random matrix
theory~\cite{rmbs,laloux99,plerou} also show the existence of correlations that
are present through a wide range of time scales from
30~mins~\cite{plerou} up to 1~day~\cite{rmbs,laloux99}. These
studies~\cite{rmbs,laloux99,plerou} show that the largest eigenvalue of the
cross-correlation matrix corresponds to correlations that pervade the
entire market, and a few other large eigenvalues correspond to
clusters of companies that are correlated amongst each other.

\section{Acknowledgments}

We thank J.-P. Bouchaud, M. Barth\'elemy, S.~V. Buldyrev, P. Cizeau,
X. Gabaix, I. Grosse, S. Havlin, K. Illinski, C.~King, C.-K. Peng,
B.~Rosenow, D. Sornette, D. Stauffer, S. Solomon, J.~Voit, and
especially R.~N. Mantegna for stimulating discussions and helpful
suggestions.  We thank X.~Gabaix, C.~King, J.~Stein, and especially
T.~Lim for help with obtaining the data.  We are also very
grateful to L.~Giannitrapani of the SCV at Boston University for her
generous help in allocating the necessary computer resources, and to
R.~Tompolski for his help throughout this work. MM thanks DFG and LANA
thanks FCT/Portugal for financial support. The Center for Polymer
Studies is supported by NSF.

%%%%%%%%%%%%%%%%%%%%%%%%%%%%%%%%%%%%%%%%%%%%%%%%
\appendix
%%%%%%%%%%%%%%%%%%%%%%%%%%%%%%%%%%%%%%%%%%%%%%%%

%%%%%%%%%%%%%%%%%%%%%%%%%%%%% SECTION
\section{Dependence of volatility on size}
%%%%%%%%%%%%%%%%%%%%%%%%%%%%%%%%%%%%%%%%%%%%%%%%

\begin{figure}
\narrowtext
\centerline{
\epsfysize=0.5\columnwidth{\rotate[r]{\epsfbox{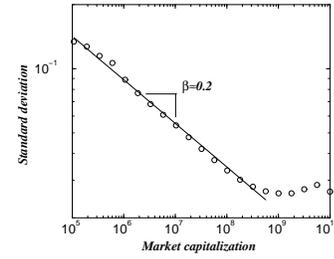}}}
}
\caption{ Log-log plot of the standard deviation of the distribution
of returns as a function of market capitalization for $\Delta t =
1$~day. Our preliminary data suggest a power-law dependence with
exponent $\beta \approx 0.2$.  This value is not unlike what was
observed for the firm sales ($\beta \approx
1/6$)~\protect\cite{Stanley96}, GDP of countries ($\beta \approx
1/6$)~\protect\cite{Lee98}, and research budgets ($\beta \approx
1/4$)~\protect\cite{pagms99}. For large values of market
capitalization, this power-law is followed by a ``flat'' region.}
\label{varsize}
\end{figure}

We find that the average volatility for each bin, $v_S(\Delta t)$
shows an interesting dependence on the market capitalization. In
Fig.~\ref{varsize}, we plot the standard deviation as a function of
size on a log-log scale for $\Delta t = 1\,$ day. We find a power-law
dependence of the standard deviation of the returns on the market
capitalization, with exponent $\beta \approx 0.2$ very similar to the
values reported for the annual sales of
firms\cite{Stanley96,Lee98,Amaral98}, the GDP of countries\cite{Lee98}
and the university research budgets\cite{pagms99}.  For larger time
scales the exponent gradually decreases, approaching the value $\beta
\approx 0.09$ for $\Delta t$= 1000$\,$days.

%In the context of scaling of the distribution of returns for various
%time scales, $\Delta t$, it is meaningful to analyze the dependence of
%the average volatility, $v_S(\Delta t)$, of each bin of average market
%capitalization $S$ on $\Delta t$. In Fig.~\ref{varsize}(b), we plot
%the standard deviation scaled by $(\Delta t)^{1/2}$, in order to see
%whether the returns for individual companies remain uncorrelated for
%large time scales.  We find that the individual companies with large
%market capitalizations scale with increasing $\Delta t$ as $(\Delta
%t)^{\gamma}$, where $\gamma \approx 0.5$, reflecting the lack of
%correlation in price changes, but companies with small values of
%market capitalizations scale with $\gamma \leq 0.5 $, which indicates
%the possibility of anti-correlation in the returns of individual
%companies with small values of market capitalization.

%%%%%%%%%%%%%%%%%%%%%%%%%%%%% REFERENCES
 
%%%%%%%%%%%%%%%%%%%%%%%%%%%%%%%%%%%%%%%%%%%%%%%%

%%%%%%%%%%%%%%%%%%%%%%%%%%%%%%%% TABLE 1
\begin{table}[hbt]
\narrowtext 
\caption{ The values of the exponent $\alpha$ for different time
scales $\Delta t$ obtained by (a) power-law regression fit to the
cumulative distribution , and (b) Hill estimator.  The non-daggered
values are computed using the TAQ database, which contains tic-data,
while the daggered values are computed using the CRSP database, which
contains records with $\Delta t = 1$~day and $\Delta t = 1$~month
sampling.  Note that we use the conversion 1~day = 390~min and
1~month = 22~days. \protect\vspace*{0.7cm} }

\begin{tabular}{cllll}
\multicolumn{1}{c}{$\Delta t~(min)$}	
	&\multicolumn{2}{c}{Power law fit} 
	& \multicolumn{2}{c}{Hill estimator}\\ \cline{2-5} 
		&Positive	&Negative 	&Positive  	&Negative \\
\hline
5      &$3.10 \pm 0.03$ &$2.84 \pm 0.12$  &$2.84 \pm 0.12$ &$2.73 \pm 0.13$\\
10     &$3.32 \pm 0.08$ &$2.89 \pm 0.13$  &$3.14 \pm 0.10$ &$2.68 \pm 0.14$\\
20     &$3.25 \pm 0.08$ &$2.75 \pm 0.10$  &$3.32 \pm 0.18$ &$2.41 \pm 0.10$\\
40     &$3.28 \pm 0.08$ &$2.61 \pm 0.10$  &$3.39 \pm 0.16$ &$2.62 \pm 0.11$\\
80     &$3.50 \pm 0.13$ &$2.49 \pm 0.11$  &$3.65 \pm 0.26$ &$2.53 \pm 0.14$\\
160    &$3.47 \pm 0.08$ &$2.42 \pm 0.09$  &$2.9 \pm 0.4$   &$2.53 \pm 0.17$\\
320    &$3.60 \pm 0.10$ &$2.54 \pm 0.10$  &$3.32 \pm 0.08$ &$3.19 \pm 0.05$\\
390$^{\dagger}$   &$2.96 \pm 0.09$ &$2.70 \pm 0.10$ &$3.05 \pm 0.13$ &$2.95 \pm 0.15$\\
780$^{\dagger}$   &$3.09 \pm 0.03$ &$2.62 \pm 0.04$ &$3.11 \pm 0.09$ &$2.90 \pm 0.12$\\
1560$^{\dagger}$  &$3.18 \pm 0.05$ &$2.75 \pm 0.09$ &$3.20 \pm 0.08$ &$2.90 \pm 0.10$\\
3120$^{\dagger}$  &$3.31 \pm 0.08$ &$2.71 \pm 0.03$ &$3.25  \pm 0.06$  &$2.94 \pm 0.09$\\
6240$^{\dagger}$  &$3.43 \pm 0.04$ &$2.74 \pm 0.12$ &$3.35  \pm 0.04$  &$2.93 \pm 0.07$\\
12480$^{\dagger}$ &$3.73 \pm 0.04$ &$2.63 \pm 0.06$ &$3.54  \pm 0.05$  &$2.93  \pm 0.08$ \\
24960$^{\dagger}$ &$3.98 \pm 0.09$ &$2.78 \pm 0.07$ &$3.89  \pm 0.09$  &$3.00  \pm 0.10$ \\
49920$^{\dagger}$ &$4.24 \pm 0.09$ &$2.84 \pm 0.07$ &$4.52  \pm 0.22$  &$3.10  \pm 0.18$ \\
99840$^{\dagger}$ &$5.06 \pm 0.07$ &$3.01 \pm 0.07$ &$4.5  \pm 0.6$  &$2.92  \pm 0.19$ \\
199680$^{\dagger}$ &$5.24 \pm 0.12$ &$3.32 \pm 0.06$ &$5.6  \pm 1.0$  &$3.14  \pm 0.13$ \\
399360$^{\dagger}$ &$6.43 \pm 0.29$ &$3.48 \pm 0.07$ &$5.11  \pm 0.03$  &$3.45  \pm 0.02$ \\
\end{tabular}
\label{alpha.exp}
\end{table}
 
%%%%%%%%%%%%%%%%%%%%%%%%%%%%%%%%%%%%%%%%%%%%%%%%

%%%%%%%%%%%%%%%%%%%%%%%%%%%%%%%%%%%%%% FIGURE 1

%%%%%%%%%%%%%%%%%%%%%%%%%%%%%%%%%%%% FIGURE 2

%%%%%%%%%%%%%%%%%%%%%%%%%%%%%%%%%%%% FIGURE 3

%%%%%%%%%%%%%%%%%%%%%%%%%%%%%%%%%%%% FIGURE 4 

%%%%%%%%%%%%%%%%%%%%%%%%%%%%%%%%%%%% FIGURE 5 

%%%%%%%%%%%%%%%%%%%%%%%%%%%%%%%%%%%% FIGURE 6 

%%%%%%%%%%%%%%%%%%%%%%%%%%%%%%%%%%%% FIGURE 7 

%%%%%%%%%%%%%%%%%%%%%%%%%%%%%%%%%%%% FIGURE 8 

%%%%%%%%%%%%%%%%%%%%%%%%%%%%%%%%%%%% FIGURE 9 

%%%%%%%%%%%%%%%%% FIGURE 10

%%%%%%%%%%%%%%%%%%%%%%%%%%%%%%%%%%%% FIGURE 11

\end{multicols}


\begin{references}
\bibitem{work_hrs} The New York Stock Exchange is open from Monday
through Friday 9:30 a.m. to 4:00 p.m. The time runs over the working
hours only. Nights, week-ends, and holidays are removed.
\bibitem{stime} The trading frequency increases, on average, with market
capitalization. For the largest companies there are several trades that
occur within each 5~min interval.  On the other hand, for the smallest
companies we consider, the typical time between trades is of the order
of 5~min.
\bibitem{errorbars} The errors on the exponent estimates are the errors
given by the regression fits to the cumulative distribution.
\bibitem{sp_diff} $S_{N=100}(t)$ and $S_{N=500}(t)$ are not exactly
identical to the S\&P100 and S\&P 500 indices, because the latter sums
the market values of the companies representing major industries at
time $t$, which are not necessarily the largest, while the former sums
over a {\it fixed\/} set of companies which are the largest in market
values on January 3, 1994. However, the difference between the two is
negligible for the period studied.
\bibitem{weights} If the weighted sum $G_{SP500}(t,\Delta t)$, in the
presence of the weights $w_i \propto S_i $, were to be dominated by
just a few of the companies ---the ones with the largest $S_i$---,
then the collapse would be trivial.  To shown that it is not so, we
compute the cumulative distribution for the returns of $X(t,\Delta t)
\equiv \sum_{i=1}^N G_i(t,\Delta t)$, which we find to coincide with
the cumulative distribution for $G_{SP500}(t,\Delta t)$.
\end{references}
\end{document}